\documentclass[fleqn,usenatbib]{mnras}

\usepackage{newtxtext,newtxmath,bm}
\usepackage[T1]{fontenc}

\DeclareRobustCommand{\VAN}[3]{#2}
\let\VANthebibliography\thebibliography
\def\thebibliography{\DeclareRobustCommand{\VAN}[3]{##3}\VANthebibliography}

\usepackage{graphicx}	% Including figure files
\usepackage{amsmath}	% Advanced maths commands
\usepackage{gensymb} 	% General symbols
\usepackage{orcidlink}
\usepackage{float}
\usepackage{comment}

\def\ddr{\frac{\partial}{\partial r}}  % Partial with respect to R
\def\vecL{ \bm{L} }                  % Angular momentum vector
\def\vecl{\bm{l}}                    % Unit angular momentum vector
\def\rin{ r_{\rm{in}} }                  % Inner radius
\def\rout{ r_{\rm{out}} }                % Outer radius
\def\abin{a_{\rm b}}                 % Binary separation
\def\ebin{e_{\rm b}}                 % Binary eccentricity

\title[Warped Circumbinary Discs]{Warps and Breaks in Circumbinary Discs}

\author[I. Rabago et al]
{
Ian Rabago$^{1,2}$ \orcidlink{0000-0001-5008-2794},
Zhaohuan Zhu$^{1,2}$ \orcidlink{0000-0003-3616-6822},
Stephen Lubow$^{3}$ \orcidlink{0000-0002-4636-7348}, and
Rebecca G. Martin$^{1,2}$ \orcidlink{0000-0003-2401-7168}
\\
$^{1}$Department of Physics and Astronomy, University of Nevada, Las Vegas 4505 S. Maryland Parkway Las Vegas, NV 89154, USA\\
$^{2}$Nevada Center for Astrophysics, University of Nevada, Las Vegas 4505 S. Maryland Parkway Las Vegas, NV 89154, USA\\
$^{3}$Space Telescope Science Institute, Baltimore, MD 21218\\
}

\date{Accepted XXX. Received YYY; in original form ZZZ}

\pubyear{20XX}

\begin{document}
\label{firstpage}
\pagerange{\pageref{firstpage}--\pageref{lastpage}}
\maketitle

\begin{abstract}
Disc warping, and possibly disc breaking, has been observed in protoplanetary discs around both single and multiple stars.  Large warps can break the disc, producing multiple observational signatures.  In this work, we use comparisons of disc timescales to derive updated formulae for disc breaking, with better predictions as to when and where a disc is expected to break and how many breaks could occur.  Disc breaking is more likely for discs with small inner cavities, cooler temperatures, and steeper power-law profiles, such that thin, polar-aligning discs are more likely to break.  We test our analytic formulae using 3D grid-based simulations of protoplanetary discs warped by the gravitational torque of an inner binary.  We reproduce the expected warp behaviors in different viscosity regimes and observe disc breaking at locations in agreement with our derived equations.  As our simulations only show disc breaking when disc viscosity is low, we also consider a viscous criterion for disc breaking, where rapid alignment to the precession vector can prevent a break by reducing the maximum misalignment between neighboring rings.  We apply these results to the GW Orionis circumtriple disc, and find that the precession induced from the central stars can break the disc if it is relatively thin.  We expect repeated or multiple disc breaking to occur for discs with sufficiently steep power law profiles.  We simulate a polar-aligning disc around an eccentric binary with steep power-law profiles, and observe two separate breaking events at locations in rough agreement with our analytical predictions.

\end{abstract}


%%%%%%%%%%%%%%%%%%%%%%%%%%%%%%%%%%%%%%%%%%%%%%%%%%

%%%%%%%%%%%%%%%%% INTRODUCTION %%%%%%%%%%%%%%%%%%%

%%%%%%%%%%%%%%%%%%%%%%%%%%%%%%%%%%%%%%%%%%%%%%%%%%
\section{Introduction}

Many astrophysical discs are observed to not lie entirely within a single plane.  These discs are generally referred to as ``warped'' discs, changing their orientation in three-dimensional space with radius.  Warped discs have been found at all different astrophysical scales through direct observation, from planetary rings to the discs of galaxies \citep{Burke1957,Kerr1957,Shu1983,Burrows1995}.   Many other warps have been inferred from their effects on their nearby surroundings, including maser emission in the vicinity of supermassive black holes \citep{Miyoshi1995,Martin2008}, to explain long-term periodicity in close X-ray binaries \citep{Katz1973,Petterson1991,Scott2000}, and shadows cast in protoplanetary discs \citep{Marino2015,Debes2017}.  The ubiquity of such a structure warrants careful study of its behavior and evolution.

The warping of discs is a well-studied phenomenon, and a large amount of work has gone into deriving the analytic behavior of the evolution and testing this behavior via numerical methods
(see \citealt{Papaloizou1983,Pringle1992,Papaloizou1995,Ogilvie1999,Lodato2007,Lodato2010}, as well as \citealt{Nixon2016} for an overall review).  In a warped disc, the angular momentum vector of the disc $\bm{L}$ changes with radius, $r$. The overall evolution of a warped disc can be effectively described with two different behaviors, depending on the relative sizes of the disc aspect ratio of scale  height to radius, $h/r$, to its $\alpha$-viscosity coefficient \citep{Shakura1973}.  Thin, high-viscosity discs with $\alpha > h/r$ are said to be in the \emph{diffusive} or \emph{viscous} regime, where the warp flattens diffusively on timescales inversely proportional to a second viscosity coefficient $\nu_2$.  Thicker, low viscosity discs with $\alpha < h/r$, are instead considered to be in the \emph{bending wave} or \emph{wave-like} regime, in which the warp launches a wave which propagates through the disc. 

In the diffusive regime, the evolution of a warped disc may be described as a diffusion equation in $\vecL$ with terms proportional to the standard viscosity $\nu_1$ and a warp viscosity $\nu_2$.  Terms including $\nu_1$ describe the standard evolution of a flat disc (i.e. \citealt{LyndenBellPringle1974}), while terms including $\nu_2$ describe the evolution of the warp in the disc \citep{Pringle1992}.  There are also additional terms present including a third viscosity term $\nu_3$, which describe an induced precession from the warp as neighboring annuli induce a torque perpendicular to the direction of the warp \citep{Ogilvie1999}.  Subsequent works have also derived evolution equations in the bending wave regime in which the effects of pressure dominate over viscosity  \citep{Papaloizou1995, Demianski1997,Lubow2000}, as well as more general sets of equations that encompass both regimes \citep{Martin2019,Dullemond2022}.

The relative sizes of the viscosity coefficients $\nu_1$, $\nu_2$, and $\nu_3$ also depend on the warp amplitude, characterized by the dimensionless parameter $\psi$ which is defined as
\begin{equation}
    \psi = r \left| \frac{\partial \bm{l}(r) }{\partial r} \right|.
    \label{eq:psi}
\end{equation}
The dependence of the coefficients on $\psi$ for large, non-linear warps and the disc viscosity $\alpha$ is characterized in \cite{Ogilvie1999}.

When warps become very large, the disc may be unable to communicate across the warp effectively. The warp may become unstable, and disc breaking can occur, splitting the disc into separate rings.  This phenomenon has been studied analytically \citep{Dogan2018,Raj2021}, and previously observed in simulations of discs in binary systems and around single and binary black holes \citep{Larwood1996,Nixon2012,Nixon2013,Nealon2015}. Disc breaking is also thought to have occurred in some discs as a possible explanation for the observed geometry or accretion kinematics  \citep{Casassus2015,Facchini2018,Zhu2019,Kraus2020,Nealon2022}.  However, the exact conditions that lead a disc to break are still unclear, as well as where the location of the break (known as the breaking radius) will occur.

In this paper, we attempt to improve the current known methods of determining a disc breaking event by studying the evolution of a disc under the influence of an external precession torque.  In our context, we consider a misaligned disc around a binary star system, analogous to protoplanetary discs forming in young star clusters.  We will see that under the influence of the binary, the disc can develop warps and breaks depending on the internal conditions of the disc.

The previous studies of disc breaking in binaries have been largely based
on the use of the SPH code. With this Lagrangian code the resolution is determined by the particle density and is therefore a concern near disc breaks. The process of the initial breaking event, as well as the subsequent evolution of the break and interaction between the misaligned inner and outer discs, involve sparse regions which may be poorly characterized in an SPH simulation unless large numbers of particles are used.
In this paper we apply an Eulerian code whose resolution is instead limited by the grid size. In addition, we take advantage of the fact that disc breaking occurs more easily for circumbinary discs that are evolving to a polar orientation, compared to evolving to a coplanar orientation, around eccentric orbit binaries.

Our paper is organized as follows.  In Section \ref{sec:timescales}, we discuss the timescales relevant to disc breaking and describe our new method of calculating the location of a break.  Section \ref{sec:methods} outlines the numerical methods we use to test our new method for calculating the breaking radius.  We present our results in Section \ref{sec:results}.  We discuss these results in Section \ref{sec:discussion}, and conclude in Section \ref{sec:conclusion}.

%%%%%%%%%%%%%%%%%%%%%%%%%%%%%%%%%%%%%%%%%%%%%%%%%%

%%%%%%%%%%%%%%%% TIMESCALES %%%%%%%%%%%%%%%%%%%%%%

%%%%%%%%%%%%%%%%%%%%%%%%%%%%%%%%%%%%%%%%%%%%%%%%%%

\section{Timescales of Disc Evolution}
\label{sec:timescales}

In this section, we examine the different evolutionary timescales that affect a warped disc, and consider how these timescales determine the breaking radius of the disc.  We consider the situation of a circumbinary disc surrounding a central binary.  The binary consists of two stars with masses $M_1$ and $M_2$, with total mass labeled $M_{\rm tot} = M_1 + M_2$, orbital semi-major axis $\abin$, and orbital eccentricity $\ebin$.  We choose radial power-law profiles for midplane density, temperature, and surface density, using the indices $d$, $s$, and $p$ given by  \footnote{Previous works commonly use $p$ for the midplane density index and $q$ for either the temperature or sound speed index.  There is also some inconsistency when considering the sign of the index, i.e. if the negative sign of the exponent is included in the variable (e.g. \cite{Takeuchi2002}).  We choose positive indices and our variable names to remain consistent with the variables in \cite{Lubow2018}, and to avoid confusion with the variable used for the binary mass ratio.}
\begin{equation}
    \rho(r) = \rho_0\left(\frac{r}{r_0}\right)^{-d},
    \label{eq:powerlaw_density}
\end{equation}
\begin{equation}
    T(r) = T_0\left(\frac{r}{r_0}\right)^{-s}
    \label{eq:powerlaw_temp}
\end{equation}
and
\begin{equation}
    \Sigma(r) = \Sigma_0\left(\frac{r}{r_0}\right)^{-p}.
    \label{eq:powerlaw_sd}
\end{equation}
For a steady state Keplerian disc, these power-law indices are related through the equation
\begin{equation}
    2d + s - 2p = 3
    \label{eq:diskprofiles}
\end{equation}
since $\rho = \Sigma/H$ \citep[e.g.][]{Pringle1981} and the sound speed is given by $c_{\rm s}=h \Omega \propto T^{1/2}$.

\subsection{Disc breaking in the wave-like regime}

\begin{figure}
    \centering
    \includegraphics[width=\columnwidth]{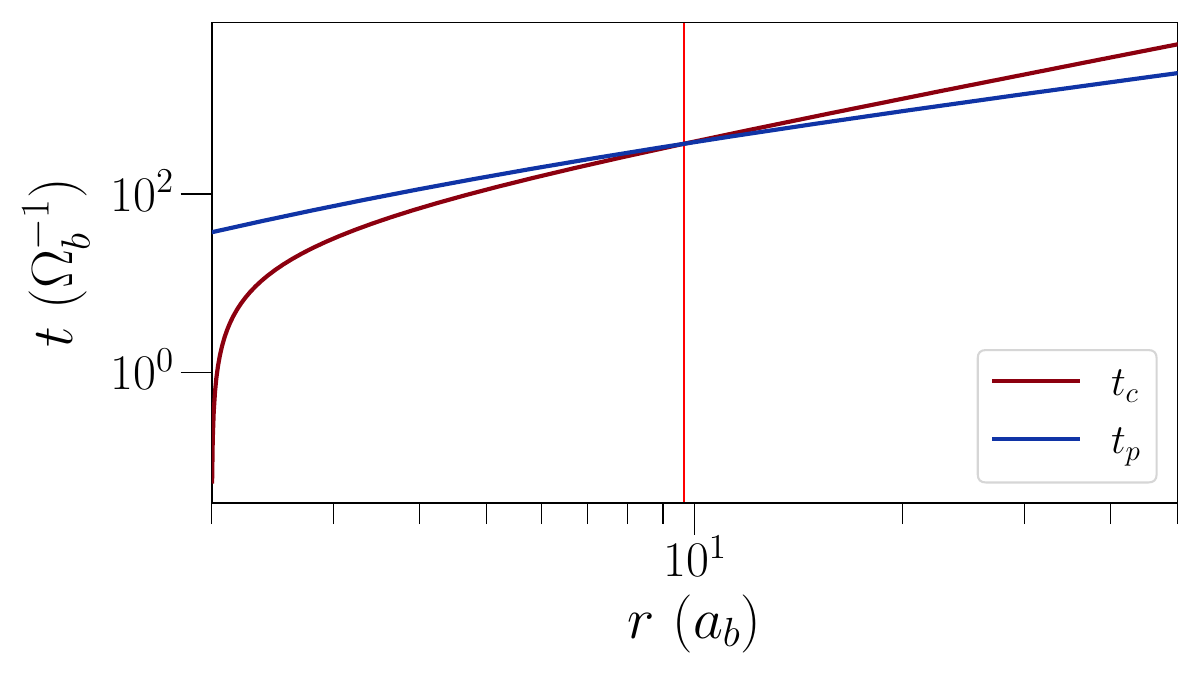}
    \caption{Disc timescales $t_{\rm p}$ and $t_{\rm c}$ as a function of disc radius on a logarithmic scale.  The red vertical line indicates the expected location of a disc breaking event based on Equations (\ref{eq:tprec_nl}) and (\ref{eq:tcomm_nl}).  For this figure, we consider a nearly coplanar disc around an equal mass binary with $e_{\rm b}=0.5$, extending from $\rin = 2a_{\rm b}$ to $\rout = 50a_{\rm b}$.  The disc has a constant scale height $h/r = 0.1$, with power law slopes of $p=1.5$ and $s = 1.0$.}
    \label{fig:tcomp}
\end{figure}

%\cite{Lubow2018} gives estimates on the time evolution of a disk under the influence of a central binary.  
The torque from the binary potential causes a ring of material to nodally precess at a frequency
\begin{equation}
    \omega_{\rm p} (r) = k \left( \frac{a_{\rm b}}{r} \right)^{7/2} \Omega_{\rm b}
    \label{eq:freqprec}
\end{equation}
\citep[e.g.][]{Farago2010},
where $\Omega_{\rm b}$ is the orbital frequency of the binary, $r$ is the radius of the ring, and $k$ is a constant which depends on the binary masses, eccentricity, and whether the disc is close to a coplanar or polar configuration \citep{Aly2015, Martin2017, Zanazzi2018, Lubow2018,Martin2018}.  For a circumbinary disc, the inner regions feel a stronger torque from the binary %precess faster 
than the outer regions. In the absence of communication between the rings of the disc, the inner rings precess faster than the outer rings. Therefore depending upon the communication timescale between the rings of the disc, the disc may become warped. %generating a warp.  

If the radial communication timescale is short compared with the precession timescale then the disc can precess like a solid body. For a disc with inner radius $\rin$  and outer radius $\rout$, the global precession timescale $t_{\rm p}$ of the disc as a solid body is given by
\begin{equation}
    t_{\rm p} = \frac{ 2(1+p) \rin^{1+p} \rout^{5/2-p} }{ |k|(5-2p)a_{\rm b}^{7/2}\Omega_{\rm b} }.
    \label{eq:tprec}
\end{equation}
The radial communication timescale in the wave-like regime $t_{\rm c}$ is given by
\begin{equation}
    t_{\rm c} = \frac{ 4 }{ (2+s)\Omega_{\rm b} h_{\rm{out}} } \left( \frac{\rout}{a_{\rm b}} \right)^{3/2}
    \label{eq:tcomm}
\end{equation}
(Eqs. 33 and 31 of \cite{Lubow2018}) where $h_{\rm out}$ is the scale height at the outer edge of the disc.

%$a_{\rm b}$ and $\Omega_{\rm b}$ are the orbital separation and orbital frequency of the binary.

We expect disc breaking to occur when the global precession timescale is shorter than the communication timescale,  $t_{\rm p} < t_{\rm c}$.  Analytic estimates of the disc breaking radius have been given in many previous works (e.g. \citealt{Nixon2013, Lubow2018}).  Combining Equations~(\ref{eq:tprec}) and~(\ref{eq:tcomm}) and solving for $r_{\rm out}$ as the breaking radius gives
\begin{equation}
    r_{\rm break} = \left[  \frac{ (p+1)(s+2) h_{\rm{out}} }{ 2 (5-2p) |k| } \left( \frac{\rin}{a_{\rm b}} \right)^{p+1} \right]^{ \frac{1}{p-1} } a_{\rm b}.
    \label{eq:rbreak}
\end{equation}
The timescales given in Equations~(\ref{eq:tprec}) and~(\ref{eq:tcomm}) assume that $\rin \ll \rout$, i.e. that the discs have a large radial extent. During breaking events, newly created discs may break off as radially thin rings, so a more complete description of these equations is required.

%\begin{equation}
%\begin{split}
%    \omega_{\rm p} = \frac{3\sqrt{5}}{4} e_{\rm b} & \sqrt{1+4e_{\rm b}^2} \frac{M_1 M_2}{M^2} \Biggl \langle \left(\frac{a_{\rm %b}}{r}\right)^{7/2} \biggr \rangle \Omega_{\rm b} \\
%    & = k \frac{\int_{\rin}^{\rout} \Sigma r^3 \Omega \left( a_{\rm b}/r \right)^{7/2}\ dr }{\int_{\rin}^{\rout} \Sigma r^3 \Omega\ dr} %\Omega_{\rm b}
%\end{split}
%\label{eq:diskfreq}
%\end{equation}

%In the second line, we gather the constants into the general constant $k$. Appendix A of \cite{Lubow2018} gives the value of $k$ as: 

%\begin{equation}
%    k = 
%    \begin{cases}
%        \frac{3}{4} \frac{M_1 M_2}{M^2} \left( 2 + 3e_{\rm b}^2\right)  & \rm{for\ a\ coplanar\ disk,}\\
%        -\frac{3}{8} \frac{M_1 M_2}{M^2} \left( 1 + 9e_{\rm b}^2\right) & \rm{for\ a\ polar\ disk.}\\
%    \end{cases}
%    \label{eq:precconst}
%\end{equation}

Here we attempt to derive a more refined estimate of $t_{\rm p}$ and $t_{\rm c}$ by revisiting these equations.  The disc nodal precession frequency is given in Equations~(16)-(17) of \cite{Lubow2018} as
\begin{equation}
    \omega_{\rm p} = k \frac{\int_{\rin}^{\rout} \Sigma r^3 \Omega \left( a_{\rm b}/r \right)^{7/2}\ dr }{\int_{\rin}^{\rout} \Sigma r^3 \Omega\ dr} \Omega_{\rm b},
\label{eq:diskfreq}
\end{equation}
where $\Omega$ is the orbital frequency and $k$ is a general constant depending on the parameters of the binary. Equation (16) of \cite{Lubow2018} and Equation (5) of \cite{Smallwood2019} give  the value of $k$ as
\begin{equation}
    k = 
    \begin{cases} -\frac{3}{4} \sqrt{1+3e_{\rm b}^2-4e_{\rm b}^4}\frac{M_1 M_2}{M_{\rm tot}^2}  & \rm{for\ a\ nearly\ coplanar\ disc,}\\
    \frac{3}{4} \sqrt{5} e_{\rm b}  \sqrt{1+4e_{\rm b}^2}
      \frac{M_1 M_2}{M_{\rm tot}^2}    & \rm{for\ a\ nearly\ polar\ disc.}\\
    \end{cases}
    \label{eq:precconst}
\end{equation}
The above expressions for $k$ are of simple analytic forms for nearly coplanar and polar orientations. The more general case is discussed in Section \ref{sec:discussion_equations}.

The fraction containing the binary masses, $M_1 M_2/M_{\rm tot}^2$, may also be written as $q/(1+q)^2$, where $q = M_2/M_1$ is the binary mass ratio.  Equation (\ref{eq:precconst}) implies that the value of $k$ is largest (and thus the precession rate quickest) for equal mass binaries.  Low eccentricity binaries produce larger $k$ values for coplanar discs, while higher eccentricity binaries give larger values for polar discs.  When considering disc breaking, larger values of $k$ reduce the precession timescale, and thus are more favorable for producing a break.  The sign of $k$ also determines the direction of precession; a positive value of $k$ signifies counterclockwise precession (in the opposite direction of disc rotation), while a negative $k$ denotes clockwise precession.

We define the disc precession time to be $t_{\rm p} = 1/\omega_{\rm p}$.  Evaluating the integrals in Equation~(\ref{eq:diskfreq}) for a disc with the surface density profile in Equation (\ref{eq:powerlaw_sd}) and Keplerian rotation profile $\Omega = \Omega_K \propto r^{-3/2}$, we find for the disc precession time
\begin{equation}
    t_{\rm p} = \frac{-2 (1+p)}{|k|(5-2p) \Omega_{\rm b}} \left( \frac{\rin}{a_{\rm b}} \right)^{7/2} \left[ \frac{\left( \rout/\rin \right)^{5/2-p} - 1}{\left( \rout/\rin \right)^{-(1+p)} - 1} \right].
    \label{eq:tprec_nl}
\end{equation}
In the limit that $r_{\rm out} \gg r_{\rm in}$, this tends to Equation~(\ref{eq:tprec}).  

To calculate the communication time of a warp across the disc, we consider the travel speed of bending waves starting from the inner edge.  These waves travel at half the local sound speed \citep{Papaloizou1995}.
%, so we calculate $t_{\rm c}= \int_{\rin}^{\rout} \frac{2}{c_s (r)} dr$.  
Assuming a radial power-law profile for the disc temperature as in Equation (\ref{eq:powerlaw_temp}), we calculate the radial communication time as
\begin{equation}
    t_{\rm c} = \int_{\rin}^{\rout} \frac{2}{c_{\rm s} (r)} dr = \frac{4}{s+2} \left[ \rout^{s/2+1} - \rin^{s/2+1} \right].
    \label{eq:tcomm_nl}
\end{equation}
An example of these timescales are shown as a function of disc radius in Figure~\ref{fig:tcomp}.  For geometrically thin rings $(\rin \approx \rout)$, Equations~(\ref{eq:tprec_nl}) and~(\ref{eq:tcomm_nl}) reproduce the correct limiting behavior $(t_{\rm p} \approx 1/\omega_{\rm p}$ and $t_{\rm c} \approx 0)$, which Equations~(\ref{eq:tprec}) and~(\ref{eq:tcomm}) do not.

%Note that, for geometrically thin rings $(r_{\rm in} \approx r_{\rm out})$, $t_{\rm p} \approx 1/\omega_{\rm p} (R)$ and $t_{\rm c} \approx 0$, the latter of which does not occur in Eq. \ref{eq:tcomm}.  When the two curves cross, $t_{\rm p} < t_{\rm c}$ and a disk break occurs.

Using Equations~(\ref{eq:tprec_nl}) and~(\ref{eq:tcomm_nl}), we can derive an updated location for the breaking radius.  In regions where $t_{\rm p} > t_{\rm c}$, a precession induced warp can be communicated via bending waves and allow the disc to precess rigidly.  
Taking the breaking radius $r_{\rm break}$ to be the value of $r_{\rm out}$ for which $t_{\rm p}= t_{\rm c}$; for $r_{\rm out} > r_{\rm break}$ we have that $t_{\rm p} < t_{\rm c}$ and the disc will precess faster than it is able to communicate the precession to its outer edge, causing the disc to break.  This criterion approximately says that the breaking radius occurs at the distance that a bending wave can travel over a disc precession timescale, starting from the disc inner edge.  
% (at speed $c_{\rm s}/2$)

Whether the curves of $t_{\rm p}$ and $t_{\rm c}$ intersect is determined by the disc \emph{geometry} $(\rin, \rout, h/r)$, disc \emph{structure} (power-law slopes $d$, $s$, and $p$), and inner binary arrangement ($q$, $a_{\rm b}$ , and $e_{\rm b}$, which determine the constant $k$).  The slope of Equation~(\ref{eq:tprec_nl}) is determined by the surface density power-law slope $p$ and the ratio $\rout/\rin$, whereas the slope of Equation~(\ref{eq:tcomm_nl}) is determined by the temperature power-law slope $s$.  The location of the inner radius, $\rin$, sets the fastest precession rate of the entire disc and the initial height of the $t_p$ curve, and is a strong factor in determining whether or not a disc will break.  Because of this sensitive dependence on $\rin$, nearly polar aligned discs are more likely to have breaks (Section \ref{sec:methods_polar} and \ref{sec:methods_gwori}), since the reduced strength of the binary torque allows them to maintain smaller inner cavities \citep{Miranda2015,Franchini2019} and therefore faster precession rates.
In addition, for fixed binary and disc parameters, the precession rate normalized by binary frequency of a nearly polar disc is faster than for a nearly coplanar disc
for binary eccentricity $e_{\rm b} > 0.41$, as follows from Equation (\ref{eq:precconst}).

Once the disc breaks into separate rings, the outer disc will have its timescales reset as its inner radius has now moved to the location of the break.  These timescales can grow again starting from the inner edge of the outer disc and, depending on the conditions of the outer disc, they may intersect again and cause the outer disc to break into yet another set of rings. Thus, being applied repeatedly, our breaking condition could predict the breaking radii of multiple breaks. In Section \ref{sec:timescales_multibreak}, we explore the possibility of multiple disc breaking in more detail, examining which disc parameters are required to cause multiple disc breaking.

\begin{figure}
    \centering
    \includegraphics[width=\columnwidth]{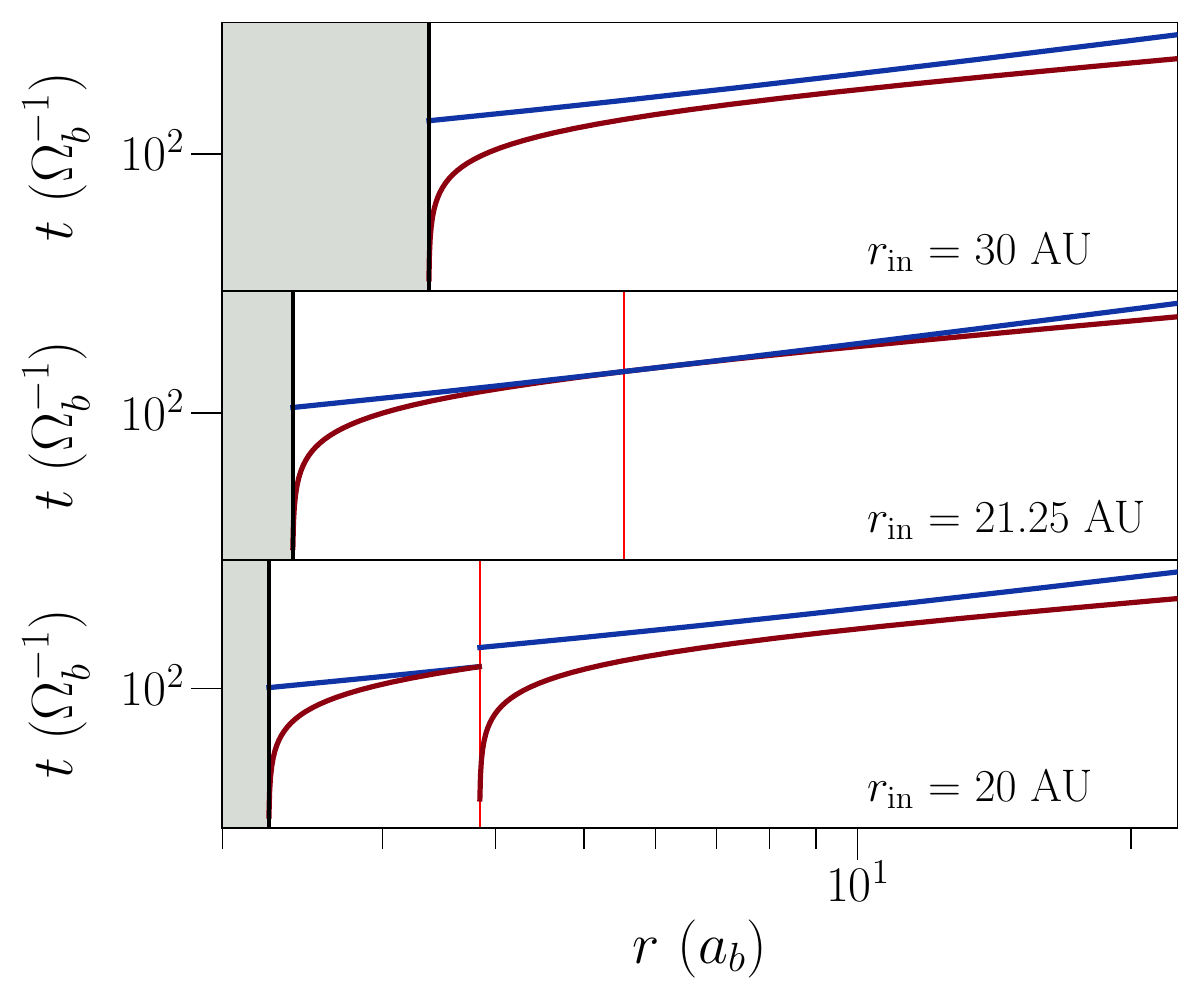}
    \caption{Predicted disc evolution of the \protect\cite{Kraus2020} parameters, as inferred from our timescale equations. In each panel the lines are the same as Fig.~\ref{fig:tcomp}. The shaded region indicates the inner cavity of the disc. \emph{Top:} Early on, the inner edge of the disc is far enough out that a break does not occur.  \emph{Middle:} As the disc evolves, the inner edge drifts inwards until $\rin = 21.25\,$au, where a break spontaneously occurs in the disc at $49 \,$au.  \emph{Bottom:} As the inner edge continues to drift inwards, so does the expected breaking radius.  Once the inner disc reaches an inner radius of 20 au, the expected breaking radius has moved to a radius of roughly 35 au. }
    \label{fig:breakevo_Kraus}
\end{figure}

\subsection{A Criterion for Viscous Disc Breaking}
\label{sec:timescales_visc}

\begin{figure*}
    \centering
    \includegraphics[width=0.95\textwidth]{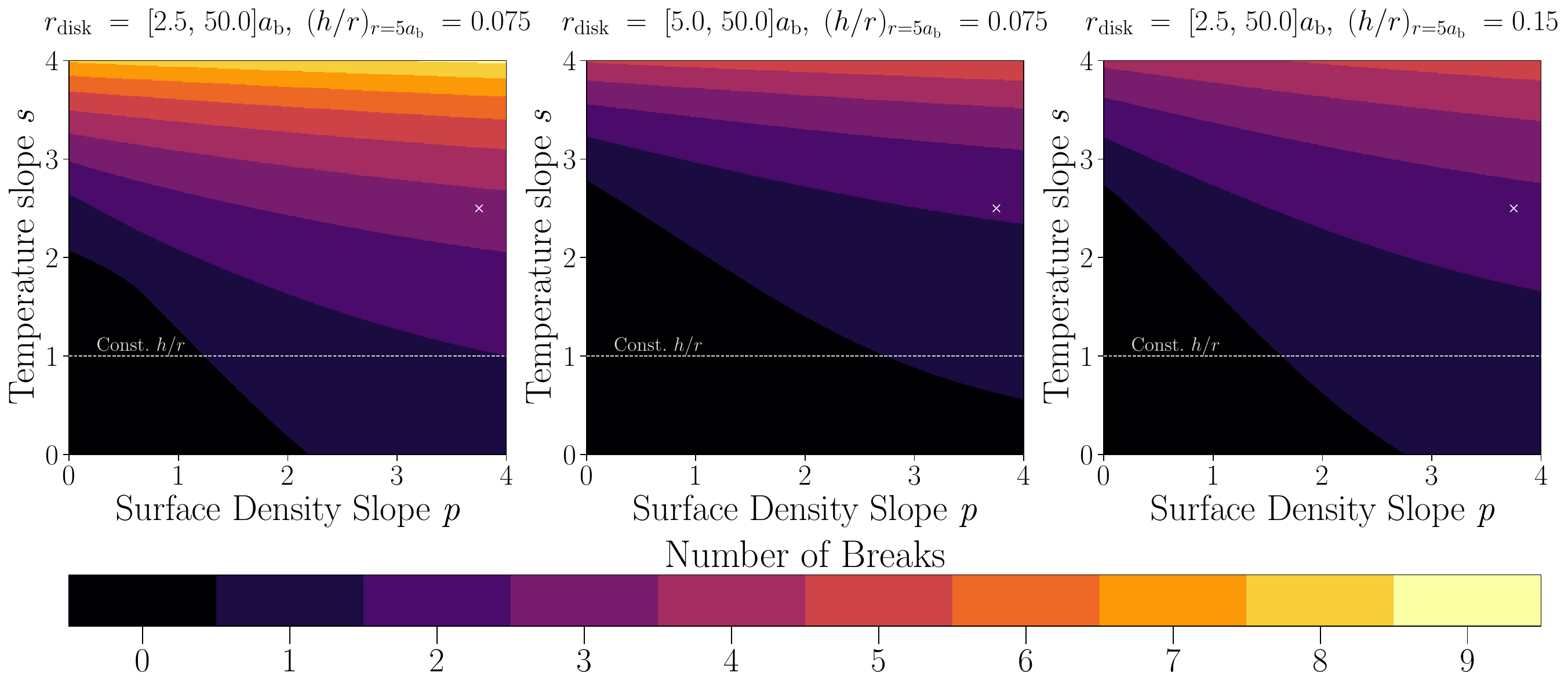}
    \caption{Predicted number of breaks for a disc with a given initial geometry, shown as a function of the surface density and temperature power-law exponents $p$ and $s$.  The radial extent of the disc is given as $r_{\rm disc} = \left[\rin,\rout\right]$ and the initial scale height is given at a distance of $r = 5a_{\rm b}$.  The central binary is an equal mass binary with eccentricity $e_{\rm b} = 0.5$.  \emph{Left:} Initial inner radius and scale height.  \emph{Middle:} Effect of a larger inner disc radius. \emph{Right:} Effect of a larger disc aspect ratio. The `$\times$' in each panel marks the disc parameters used in our multiple break simulation (Section \ref{sec:methods_multi}.)  Disc inclination is not considered as a factor in these plots, but the effect is expected to be small as long as the disc is far from the critical inclination - see Section \ref{sec:discussion_equations} for details. }
    \label{fig:breaksearch}
\end{figure*}

Discs align towards their precession vector ($\vecL_{\rm b}$ for coplanar orientations, or ${\bm e}_{\rm b}$ for polar orientations) in both the wave-like and viscous regimes on a timescale that is roughly inversely proportional to $\alpha$ (\cite{King2013}, \citealt{Lubow2018}, Eq.29)
\begin{equation}
    t_{\rm align} = \frac{(h/r)^2 \Omega_{\rm b}}{\alpha (\omega_{\rm p})^2}.
\end{equation}
If the alignment timescale is significantly faster than the precession rate,  the disc will align to the precession vector before it can precess a significant amount.  This can suppress large warps generated by azimuthal ``twisting'' motions \citep{ Raj2021}, which may prevent disc breaking even if the other disc parameters would normally allow a break to occur.  We can express this relation by finding the ratio of $t_{\rm p}$ to $t_{\rm align}$

\begin{equation}
    %\tau(r) = \frac{t_{\rm align}}{t_{\rm p}} = \left( \frac{h/r}{2\pi} \right)^2 \frac{1}{\alpha} t_{\rm p} \Omega_{\rm b}.
    \tau(r) = \frac{t_{\rm align}}{t_{\rm p}} = \frac{(h/r)^2}{\alpha} t_{\rm p} \Omega_{\rm b}.
    \label{eq:tau}
\end{equation}

When this ratio is less than 1, $t_{\rm align} < t_{\rm p}$ and alignment occurs before disc precession, preventing disc breaking.  This criterion is primarily applicable to discs in the diffusive regime, although the factor of $(h/r)^2/\alpha$ suggests it may apply to some discs in the wave-like regime as well.  Equation (\ref{eq:tau}) is a simple estimate which only considers the disc scale height at one particular radius; other approaches may take into account how the scale height changes radially across the disc.  %We note that this equation is similar to Equation (13) of \cite{King2013}, which instead compares the viscous flattening timescale of the warp $t_{\nu_2}$ to the precession frequency $\omega_{\rm p}$ and considers $t_{\nu_2}\omega_{\rm p} < 1$ as the criterion to prevent large disc warping in the viscous regime. 
Various disc breaking criteria have been suggested in the past. Equation (13) of \cite{King2013} considers $t_{\nu_2}\omega_{\rm p} < 1$ as the criterion to prevent large disc warping in the viscous regime, where  $t_{\nu_2}$ is the viscous timescale associated with the $\nu_2$ viscosity. We note that in the linear regime of small warps, this criterion and the Equation (\ref{eq:tau}) 
give opposite   stability results because $(\omega_{\rm p} \, t_{\rm align})^{-1} \sim \omega_{\rm p} \, t_{\nu2}$. A short (long) alignment timescale occurs with a long (short) warp viscous timescale. But the $\nu_2$ value for small warps may not apply for cases of disc breaking.

\subsection{An Application to GW Orionis}
\label{sec:timescales_gwori}

GW Orionis is a hierarchical triple star system surrounded by a large protoplanetary disc \citep{Mathieu1991,Berger2011}.   Resolved images of the system \citep{Bi2020,Kraus2020} show a series of dust rings in differing orientations around the central star system as well as distorted velocity maps in $^{12}\rm{CO}$, suggesting that the disc is warped.  Observations indicate that the outer disc is inclined with respect to the outer binary orbital plane at an angle of roughly $i = 38^\circ$. 

The GW Orionis system has been simulated in previous works, but simulations have shown differing results as to whether the disc will break solely under the influence of the inner star system.  In their observational paper, \cite{Kraus2020} use SPH simulations to explain the warped geometry of the system.  They simulate the GW Ori disc with an initial radial range of 40 to $200\,$au, using power-law slopes of $p = 0.2$ and $s = 1.0$ and a very thin disc with $h/r = 0.02$.  The disc expands inwards until the inner radius reaches about 30au, where a thin ring breaks off from the inner edge of the disc and precesses independently, eventually aligning with the binary plane.  \cite{Kraus2020} interpret this inner ring as a potential origin of the R3 dust ring seen in observations.

\cite{Bi2020} and \cite{Smallwood2021} ran a similar suite of SPH simulations for the GW Ori system, using a disc with the same initial inner edge but a slightly thicker aspect ratio of $h/r = 0.05$ and power-law slopes of $p = 1.5$ and $s = 1.0$.  Their simulations find a disc that warps strongly but does not break through disc precession alone. They suggest that a giant planet embedded in the disc is required to carve a gap first in order for an inner ring to break off.  They also find that a disc with a smaller inner radius, $\rin = 20\rm{au}$, can spontaneously generate a break, with a breaking radius of roughly 75 au.  However, note that this is a result of the initial surface density profile which is a power law in radius with a sharp cut off at the initial disc inner edge. The disc is not in equilibrium and over time material is cleared from this region as a result of the binary torque. The set up is therefore somewhat artificial.  Despite these problems, they determine that the geometry of the disc is an important factor in determining if a disc is able to break.

By applying Equations~(\ref{eq:tprec_nl}) and~(\ref{eq:tcomm_nl}) to the GW Orionis system, it is possible to explain the appearances of both SPH simulations.  To model this triple star system with our analytic equations, we assume that the triple star orbit can be approximated by the outer binary orbit of the triple. This assumption is justified for a low binary eccentricity \citep[e.g.][]{Lepp2023,Smallwood2021}. The components of the outer binary are composed of two masses with $M_1=0.742M_{\rm tot}$ and $M_2=0.258M_{\rm tot}$ and an eccentricity of $e_{\rm b} = 0.379$.  Including the effect of the inner binary will cause apsidal precession in the orbit of the outer binary \citep{Lepp2023}, but we do not consider this effect when considering the precession of the disc.  For the circumtriple disc, we consider the two disc models with the parameters listed above.  We use the parameters from two disc models listed above, with the size of $\rin$ varying from 40au to 20au to simulate the inward drift of the disc.

For the \cite{Kraus2020} simulation, the initial conditions have $t_{\rm p} > t_{\rm c}$ everywhere, indicating a disc with no breaks.  However, the two curves approach as the inner edge of the disc drifts inward.  Once the inner edge of the disc reaches about 21au, the two curves intersect and a break spontaneously appears at $r_{\rm break} = 49 \rm{au}$, producing a thin secondary ring along the inner edge.  By the time the inner cavity has shrunk to $20 \rm{au}$, the breaking radius has also moved inwards to a distance of $r_{\rm break} = 35 \rm{au}$, a distance that is consistent with the observed location of the ring seen in the SPH simulations.  Figure \ref{fig:breakevo_Kraus} shows the evolution of $t_{\rm p}$ and $t_{\rm c}$ in the Kraus disc model.  If spontaneous disc breaking occurs in this manner, the material in the inner ring can be shuttled inward into a tight ring similar to what is seen in the GW Ori observations.

The discs used by \cite{Bi2020} and \cite{Smallwood2021} are thicker and have a higher disc temperature, creating smaller values for $t_{\rm c}$.  Although the two curves begin to intersect at nearly the same value of $\rin$, the change in the shape of the curves causes the value of $r_{\rm break}$ to start from the outer edge of the disc and sweep inwards, instead of starting in the middle like the \cite{Kraus2020} disc.  The inner cavity must shrink even more, down to $\rin$ = 15au, in order for the break to produce a similarly confined inner ring.  This model is unlikely to break the disc using the binary torque alone, making a planet-driven explanation more likely for these disc parameters.

By comparing these simulations, it is clear that the details of the simulation parameters are important in determining the location of a disc break.  Observations of the GW Ori system are currently unable to put strong constraints on the disc parameters, so the GW Ori disc may be able to break under the torque of the binary if it is closer to the \cite{Kraus2020} simulation, i.e. cooler and with lower values of $h/r$.  In Section \ref{sec:methods_gwori}, we describe our simulations for the GW Ori system. 

\subsection{Multiple Disc Breaking}
\label{sec:timescales_multibreak}

%The interaction between the different disk timescales to produce several breaking radii is shown in Figure \ref{fig:tcomp}, showing a disk with 2 disk breaks.
%After the disk breaks, the outer disk will begin to differentially precess starting from its inner edge.  If the breaking radius of this new disk is still within the outer extent of the disk, then the disk may break again.  This process of \emph{multiple disk breaking} can repeat and break the disk into multiple rings for as long as there is disk material, or until the estimates for $t_{\rm c}$ and $t_{\rm p}$ above are no longer valid.  Some SPH simulations have observed this phenomenon previously, in the context of black hole binaries and the Bardeen-Petterson effect \citep{Nixon2013,Nealon2015}.

After a disc breaks into two, the inner and outer disc will begin to precess independently.  The inner disc is guaranteed to precesss rigidly, by the conditions set by the break, but the outer disc may once again precess differentially with its own $t_{\rm c}$ and $t_{\rm p}$ starting from the breaking radius which is its new inner edge.  If the breaking radius of this new outer disc is within the radial extent of the outer disc, then the outer disc may break again.  This process of \emph{multiple disc breaking} can repeat and break the disc into multiple rings for as long as there is disc material, or until the estimates for $t_{\rm c}$ and $t_{\rm p}$ above are no longer valid.  Some SPH simulations have observed this phenomenon previously, in the context of black hole binaries and the Bardeen-Petterson effect \citep{Nixon2013,Nealon2015}.

%The location and number of disk breaks calculated during this process is determined by the given disk \emph{geometry} $(\rin, \rout, h/r)$, disk \emph{structure} (power-law slopes $d$ and $s$, which determine $p$), and inner binary arrangement ($q, a, e$, which determine $k$).
%Each panel shows the number of expected disk breaking events for a fixed disk geometry as a function of the power-law slopes $p$ and $s$ (the disk structure).
To get a better understanding of how the various disc parameters affect the number of disc breaks, in Figure \ref{fig:breaksearch} we map out the number of disc breaks for different initial values of $\rin$, $\rout$, and $(h/r)_0$ (the disc geometry) and as a function of the power-law slopes $p$ and $s$ (the disc structure).  In this figure, the disc is placed around an equal mass $(M_1 = M_2= 0.5),\ e_{\rm b}= 0.5$ central binary.  Each panel shows the number of expected disc breaking events for the given initial disc geometry as a function of $p$ and $s$.  In each panel, the horizontal line at $s = 1$ represents discs with a constant $h/r$ throughout; points below this line represent flaring discs, with $h/r$ increasing with radius.  In general, discs that are thin, have small inner cavities, and steep profiles are more likely to have multiple breaks.  

%The location of the inner radius, $\rin$, sets the fastest precession rate of the entire disc and is a strong factor in determining whether or not a disc will break.  Doubling the inner radius reduces the number of breaks in each region by roughly 1 each.  This feature explains the simulations of \cite{Smallwood2021}, who also noticed that disc breaking was dependent on the disc inner radius.  Figure \ref{fig:breaksearch} also shows that increasing the disc scale height increases the width of each "stripe", requiring a steeper temperature gradient to achieve multiple breaks.  Because of this sensitive dependence on $\rin$, polar aligned discs are more likely to have breaks (Section \ref{sec:methods_polar} and \ref{sec:methods_gwori}), since they are able to maintain smaller inner cavities \citep{Franchini2019} and faster precession rates.

The effect of changing the inner radius can be seen by comparing the first and second panels.  Doubling the inner radius reduces the number of breaks in each region by roughly 1 each.  This behavior is consistent with the disc breaking seen in the simulations of \cite{Smallwood2021}, who also noticed that disc breaking was dependent on the disc inner radius.  Figure \ref{fig:breaksearch} also shows that increasing the disc scale height increases the width of each "stripe", requiring a steeper temperature gradient to achieve multiple breaks.  For typical protoplanetary disc parameters, disc breaking will usually be limited to a single break unless the disc is extremely thin.  However, multiple breaking may still occur naturally when considering other sources of precession, such as around black holes.  Tidal interactions from the binary component can truncate the disc edges, providing limits on the values of $\rin$ and $\rout$ \citep{Larwood1996,Lubow2015,Miranda2015}.  \cite{Miranda2015} study tidal truncation of circumbinary discs due to Lindblad torques and find that for highly inclined discs, the inner radius is truncated close to the 1:3 and 1:4 commensurabilities, approximately 2.0 to 2.5 $\abin$.

% New equation for Rbreak with new timescale equations???
%This is reflected in the exponent of $r_{\rm in}$ in Eq.\ref{eq:rbreak}.
%As mentioned in previous simulations above, the location of the inner radius, $r_{\rm in}$, is a strong factor in determining whether or not a disk will break.  Doubling the inner radius reduces the number of breaks in each region by 1 each.  Increasing the disk scale height widens the width of each "stripe", requiring a steeper temperature gradient to achieve multiple breaks.

%%%%%%%%%%%%%%%%%%%%%%%%%%%%%%%%%%%%%%%%%%%%%%%%%%

%%%%%%%%%%%%%%%%%%% METHODS %%%%%%%%%%%%%%%%%%%%%%

%%%%%%%%%%%%%%%%%%%%%%%%%%%%%%%%%%%%%%%%%%%%%%%%%%
\section{Methods}
\label{sec:methods}

\begin{figure*}
    \centering
    \includegraphics[width=0.95\textwidth]{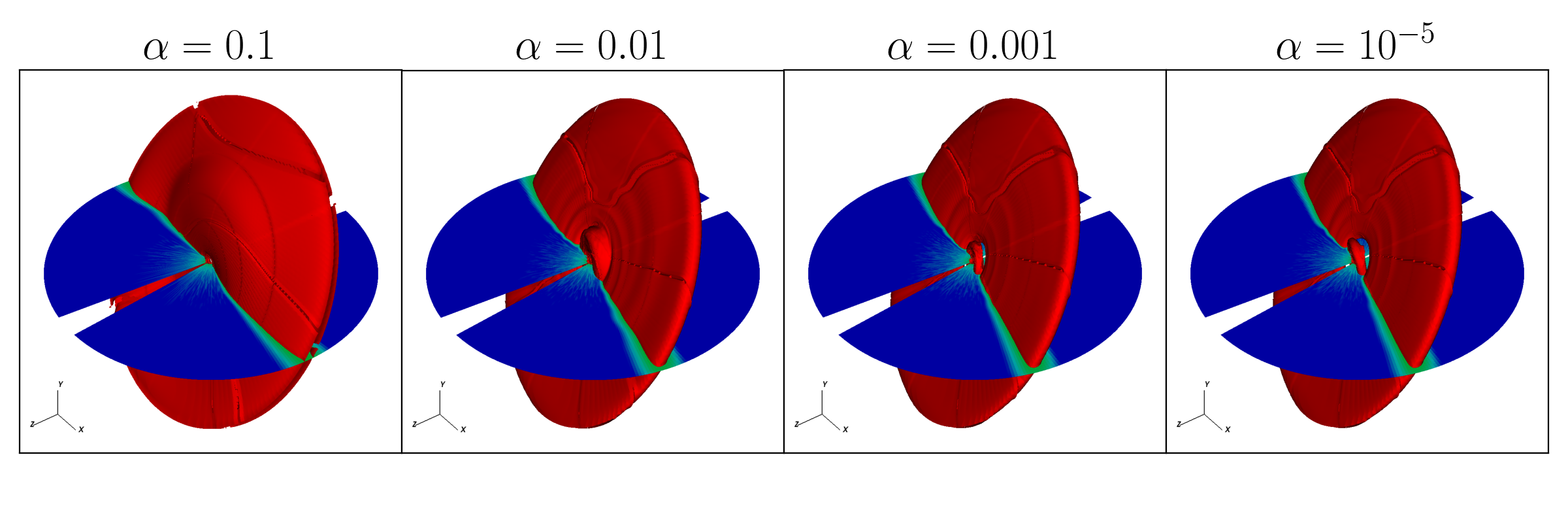}
    \caption{Density contours of the polar-aligning discs at $t=1000$ binary orbits.  Horizontal slice shows disc density, along the plane of the binary.  Discs in the $\alpha = 0.01, 0.001$, and $10^{-5}$ cases are broken; the inner disc in the $\alpha=0.01$ simulation has momentarily reconnected with the outer disc as it precesses about the binary eccentricity vector.  Please note that the radial lines visible on the density contour are a visual artifact of the rendering process and are not a part of the simulation. }
    \label{fig:polardensity}
\end{figure*}

%We use the grid-based code \textsc{ATHENA++} \citep{Stone2020} to simulate a circumbinary disk in spherical-polar coordinates $(r, \theta, \phi)$.  Our simulation domain extends from $0.3$ to  $10.0$ in $r$, $5 \degree$ to $175 \degree$ in $\theta$, and $0$ to $2\pi$ in $\phi$, and is divided into a grid of $216 \times 176 \times 384$ cells in $(r, \theta, \phi)$.

We use the grid-based code \textsc{ATHENA++} \citep{Stone2020} to simulate an inclined disc around a central binary.  To study the various effects of disc warping and breaking, we run three different sets of simulations, each with slightly different disc and binary setups.  Each set is described in the following subsections.  The disc and binary parameters used for each simulation set are listed in Table \ref{tab:simparameters}.  Below we describe features common to all of the simulations performed in this paper.

We choose spherical-polar coordinates $(r, \theta, \phi)$ for our simulations.  Our simulation domain extends from $5 \degree$ to $175 \degree$ in $\theta$, and $0$ to $2\pi$ in $\phi$, and is divided into a grid of $176 \times 384$ cells in $(\theta, \phi)$.  We use logarithmically spaced cells in the radial domain.  The extent and number of cells in the radial domain $r$ varies depending on the simulation set.  At the simulation boundaries, we set a one-way outflow boundary condition in the radial direction, and reflecting boundary conditions in the polar ($\theta$) direction.  The choice of boundary conditions in the radial direction allows bending waves in the disc to dissipate without reflection at the disc edges.  Though the boundary conditions in the polar direction are reflecting, the large extent of the polar domain and settling of the disc to the simulation midplane keeps the disc far away from these boundaries.

We initialize the density profile of the disc by numerically integrating the density at each grid cell to establish hydrostatic equilibrium according to the power-law profile in Equation (\ref{eq:powerlaw_density}), and set the vertical profile by numerically integrating the density at each grid cell to establish hydrostatic equilibrium according to the disc scale height $h = c_{\rm s}/\Omega_{\rm K}$. Here $r_0 = a_{\rm b}$, $\rho_0 = 1$ at $r_0$, $c_{\rm s}$ is the local isothermal sound speed $\sqrt{P/\rho}$ at $r$, and $\Omega_{\rm K}$ is the Keplerian orbital frequency at $r$.  The disc temperature is initialized using the power-law profile in Equation (\ref{eq:powerlaw_temp}), where  $T_0 = 1$ at $r_0$.  We truncate the edges of the disc using an exponential cutoff of the form $\exp{\left[(r-\mu_r)/\sigma_r\right]}$, where $\mu_r$ is the location of the disc edge and $\sigma_r$ the relative scale length of the cutoff.  We also use a spherically symmetric density floor with a value of $\rho_{\rm floor} = 10^{-4} \rho_0$ at $r = 1$ and a power-law slope of $d$, identical to that of the density profile.  We initialize the disc with an initial scale height of $(h/r)_0$ at $r=r_0$.  We use the orbital cooling scheme outlined in Equation (5) of \citep{Zhu2015}, using a dimensionless cooling time of $t_{\rm cool} = 0.01 \Omega_{\rm K}^{-1}$.

%We initialize the disc with an initial scale height of $(h/r)_0$ at $r=r_0$.  We use the orbital cooling scheme outlined in Equation (5) of \citep{Zhu2015}, using a dimensionless cooling time of $t_{\rm cool} = 0.01 \Omega_{\rm K}^{-1}$.  At the simulation boundaries, we set a one-way outflow boundary condition in the radial direction, and reflecting boundary conditions in the polar direction.  \textbf{The choice of boundary conditions in the radial direction allows bending waves in the disc to dissipate without reflection at the disc edges.  Though the boundary conditions in the polar direction are reflecting, the large extent of the polar domain and settling of the disc to the simulation midplane keeps the disc far away from these boundaries.}

The binary components are simulated as gravitational bodies with masses $M_1$ and $M_2$, which orbit with semi-major axis $a_{\rm b}$ and eccentricity $e_{\rm b}$.  We integrate the motion of the binary components using a 2nd-order leapfrog integrator, but we do not model the change in the orbits due to interaction with the disc material.

\begin{table}
    \centering
    \begin{tabular*}{0.95\columnwidth}{l r r r}
        Parameter & Polar (\S\ref{sec:methods_polar}) & GW Ori (\S\ref{sec:methods_gwori}) & Multibreak (\S\ref{sec:methods_multi})\\
        \hline \hline
        $M_1$               & 0.5   & 0.742 & 0.5   \\
        $M_2$               & 0.5   & 0.258 & 0.5   \\
        $\ebin$             & 0.5   & 0.379 & 0.5   \\
        \hline
        $\mu_r (\abin)$     & 2.0   & 2.7   & 2.0   \\
        $\sigma_r (\abin)$  & 0.35  & 0.35  & 0.35  \\
        $(h/r)_0$           & 0.103 & 0.075 & 0.251 \\
        $i_0 (^\circ)$      & 60    & 38    & 60    \\
        \hline
        $d$                 & 2.25  & 2.5   & 4.0   \\
        $s$                 & 1.5   & 1.0   & 2.5   \\
        $p$                 & 1.5   & 1.5   & 3.75  \\
        $\alpha$ & $10^{-1}, ^{-2}, ^{-3}, ^{-5}$ & $10^{-5}$ & $10^{-5}$
    \end{tabular*}
    \caption{Table of parameters used in the simulations.  Parameters are sorted in groups of binary arrangement (top), disc geometry (middle), and disc structure (bottom).}
    \label{tab:simparameters}
\end{table}

%$10^{-1}, 10^{-2}, 10^{-3},10^{-5}$

%-----------------------------------------------%
%--------------- POLAR DISKS -------------------%
%-----------------------------------------------%
\subsection{Polar-Aligning Warped Discs}
\label{sec:methods_polar}

Our first set of simulations studies the effects of disc warping in a polar-aligning disc around an eccentric binary.  As described in Section \ref{sec:timescales}, polar-aligning discs benefit from faster precession times due to smaller central cavities and the response from the central binary, making them ideal for studying disc breaking.  Our setup is similar to that used in \cite{Rabago2023}, which we review in brief here.  The binary components are initialized as equal mass particles $(M_1 = M_2 = 0.5)$ with $a_{\rm b} = 0.28$ and $e_{\rm b} = 0.5$, placing the binary as close to the inner radial domain as possible.  We simulate the region from $0.3=1.07 a_{\rm b}$ to $10.0=35.7 a_{\rm b}$ using 216 cells in the radial domain.  The initial scale height is set at $(h/r)_0 = 0.103$, giving the disc a scale height of $h/r = 0.075$ at $r = 3.5a_{\rm b}$. We use power-law slopes of $d = 2.25$ and $s = 1.5$ $(p = 1.5)$ and truncate the inner edge of the disc using $\mu_r = 2a_{\rm b}$ and $\sigma_r = 0.35a_{\rm b}$.  The outer edge of the disc extends to the outer radial domain.  We vary the $\alpha$-viscosity parameter between $\alpha = 10^{-1}, 10^{-2}, 10^{-3}$, and $10^{-5}$.  These values correspond to discs with $h/r < \alpha$ (diffusive regime), $h/r \sim \alpha$ (intermediate case), $h/r > \alpha$ (wave-like regime), and a case for the inviscid limit.  We place the binary along the simulation $xz$-plane and initialize the disc with a $60^\circ$ inclination to the binary orbital plane.  We run these simulations for 1000 binary orbits.

\subsection{GW Orionis}
\label{sec:methods_gwori}

We create a suite of simulations that replicates the GW Orionis system.  We use the known binary parameters from \cite{Kraus2020}, using the AB-C binary separation of $a_{\rm b} = 8.89$ au as the system scale.  Previous simulations of the GW Ori system from \cite{Smallwood2021} found that the motion of the inner AB binary provides smaller effects to the disc compared to the larger motion of the AB-C binary, so we model the GW Ori triple system as a binary to simplify calculations as in Section \ref{sec:timescales_gwori}.  Our binary model consists of the AB binary as one mass $M_{AB} = 0.742 M_{\rm tot}$ and the outer C component as the second mass $M_C = 0.258 M_{\rm tot}$, placed in an orbit with semi-major axis $a_{\rm b} = 8.89$ au and eccentricity $e_{\rm b} = 0.379$.

We simulate the region from $1.5a_{\rm b}$ to $50a_{\rm b}$, covering a range of roughly 13 au to 450 au.  We use 192 cells across the radial domain.  The disc is inclined at an angle of $38^\circ$ relative to the binary.  We use power-law exponents of $d = 2.5$ and $s = 1.0$.  This gives the disc a surface density profile of $p = 1.5$ and a constant $h/r$ throughout, which we choose to be $(h/r)_0 = 0.075$.  We truncate the inner edge of the disc at $\mu_r = 24$ au = $2.7\abin$ and use a disc viscosity of $\alpha = 10^{-5}$.    We expect this simulation to be similar to the \cite{Smallwood2021} simulation with a small inner radius, generating a break at around $10 a_{\rm b}$.  Although the disc viscosity is lower than their simulations, we expect the inner edge of the disc to be truncated at a similar radius due to the torque of the binary and, with $\alpha < h/r$, the disc remains in the wave-like regime and should break in a similar fashion.  We run these simulations for 5000 binary orbits, enough time for the inner cavity to settle and the warp to propagate through the disc. We do not attempt to reproduce the disc from \cite{Kraus2020}, due to the thin disc scale height of $h/r = 0.02$ used in their work.  Resolving this disc at the required inclination would require a large increase in the overall grid resolution, which we consider prohibitively expensive.

%-----------------------------------------------%
%--------------- MULTI BREAK -------------------%
%-----------------------------------------------%

\subsection{Multiple Breaks}
\label{sec:methods_multi}

As described in Section \ref{sec:timescales_multibreak}, the equations for $t_{\rm p}$ and $t_{\rm c}$ suggest a disc can undergo multiple breaks if the power-law slope of $t_{\rm p}$ is less than that of $t_{\rm c}$.  This can occur given the right combination of disc geometry and density profiles.  To examine this possibility, we run an additional simulation with a disc that is predicted to undergo multiple breaks by Equations~(\ref{eq:tprec}) and~(\ref{eq:tcomm}).  To maximize the chance of multiple breaks, we choose a polar-aligning disc with steep density and temperature profiles.

We modify our setup for the polar disc simulations, changing the radial domain so the inner domain remains at $\rin = 1.07a_{\rm b}$ but extending the outer radial domain to $\rout = 100a_{\rm b}$ using 272 radial cells across the radial domain.  We use power-law profiles of $d = 4.0$ and $s = 2.5$ $(p = 3.75)$.  We choose the disc scale height to be $(h/r)_{r=5a} = 0.075$.  Combined with the temperature profile, the disc scale height ranges from $(h/r)_{\rm in} = 0.251$ at the disc inner edge to $(h/r)_{\rm out} = 0.013$ at $r=50 a_{\rm b}$.  To ensure the disc is resolved as the scale height decreases, we use a single level of mesh refinement to refine the regions between $r = [8.89\abin, 52.1\abin]$ and $\theta = [50.8^\circ, 129.1^\circ]$.  The disc viscosity is set to $\alpha = 10^{-5}$, as in the nearly inviscid cases.

A disc with these parameters is predicted to undergo either two or three breaking events, depending on the exact location of the inner radius.  The location of this setup is marked on each panel of Figure \ref{fig:breaksearch} with an `$\times$'.  For the left panel, with $\rin = 2.5a_{\rm b}$, the breaks are predicted to occur at distances of 4.0, 8.6, and 29.1 $a_{\rm b}$.  We evolve this disc for 1500 binary orbits, enough time for the outer parts of the disc to evolve for a few orbital timescales.

%%%%%%%%%%%%%%%%%%%%%%%%%%%%%%%%%%%%%%%%%%%%%%%%%%

%%%%%%%%%%%%%%%%%%% RESULTS %%%%%%%%%%%%%%%%%%%%%%

%%%%%%%%%%%%%%%%%%%%%%%%%%%%%%%%%%%%%%%%%%%%%%%%%%
\section{Results}
\label{sec:results}

%-----------------------------------------------%
%--------------- POLAR DISKS -------------------%
%-----------------------------------------------%
\subsection{Disc Warping and Breaking}
\label{sec:results_polar}

\begin{figure}
    \centering
    \includegraphics[width=\columnwidth]{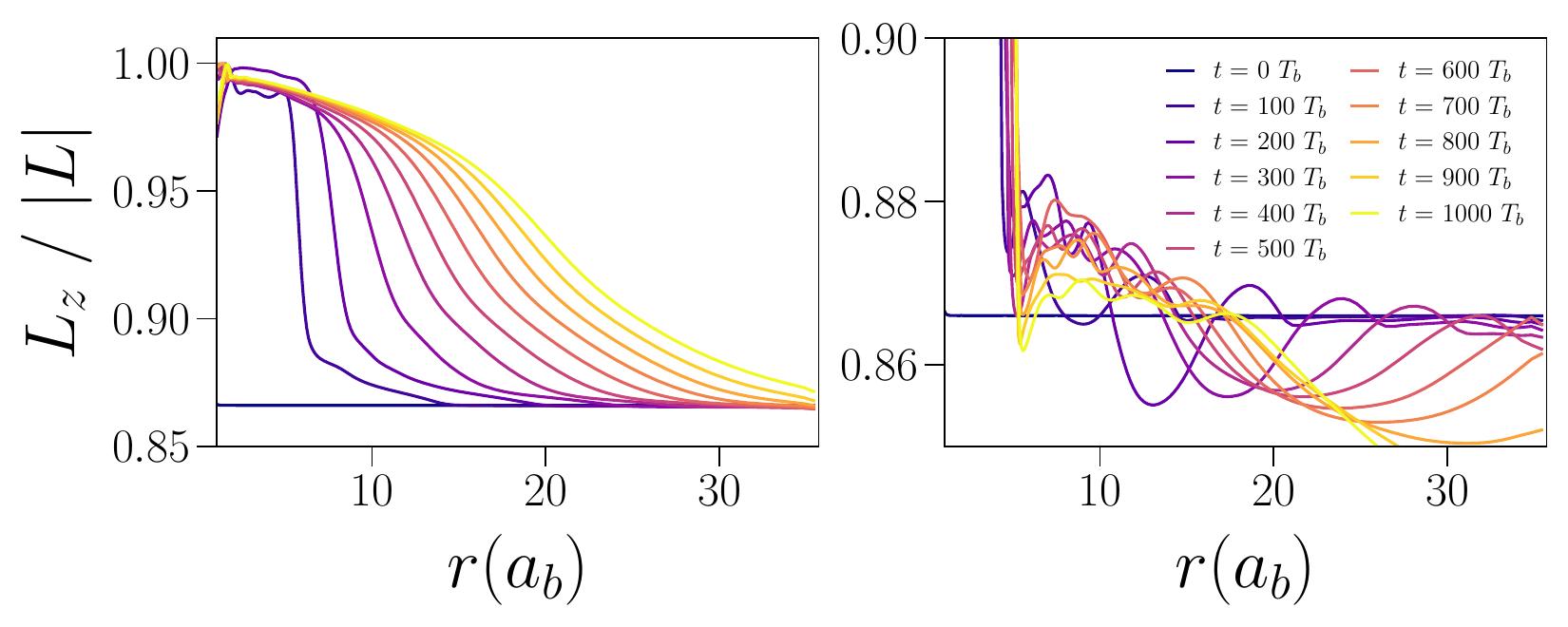}
    \caption{Evolution of the warp profile for the viscous ($\alpha = 0.1$, left) and inviscid cases ($\alpha = 10^{-5}$, right).  Each curve shows $\mathbf{l_z}$, the $z$-component of the unit angular momentum vector  (along the direction of the binary eccentricity vector) as a function of radius.  Curves are plotted every 100 binary orbits.}
    \label{fig:warpevo}
\end{figure}

\begin{figure}
    \centering
    \includegraphics[width=\columnwidth]{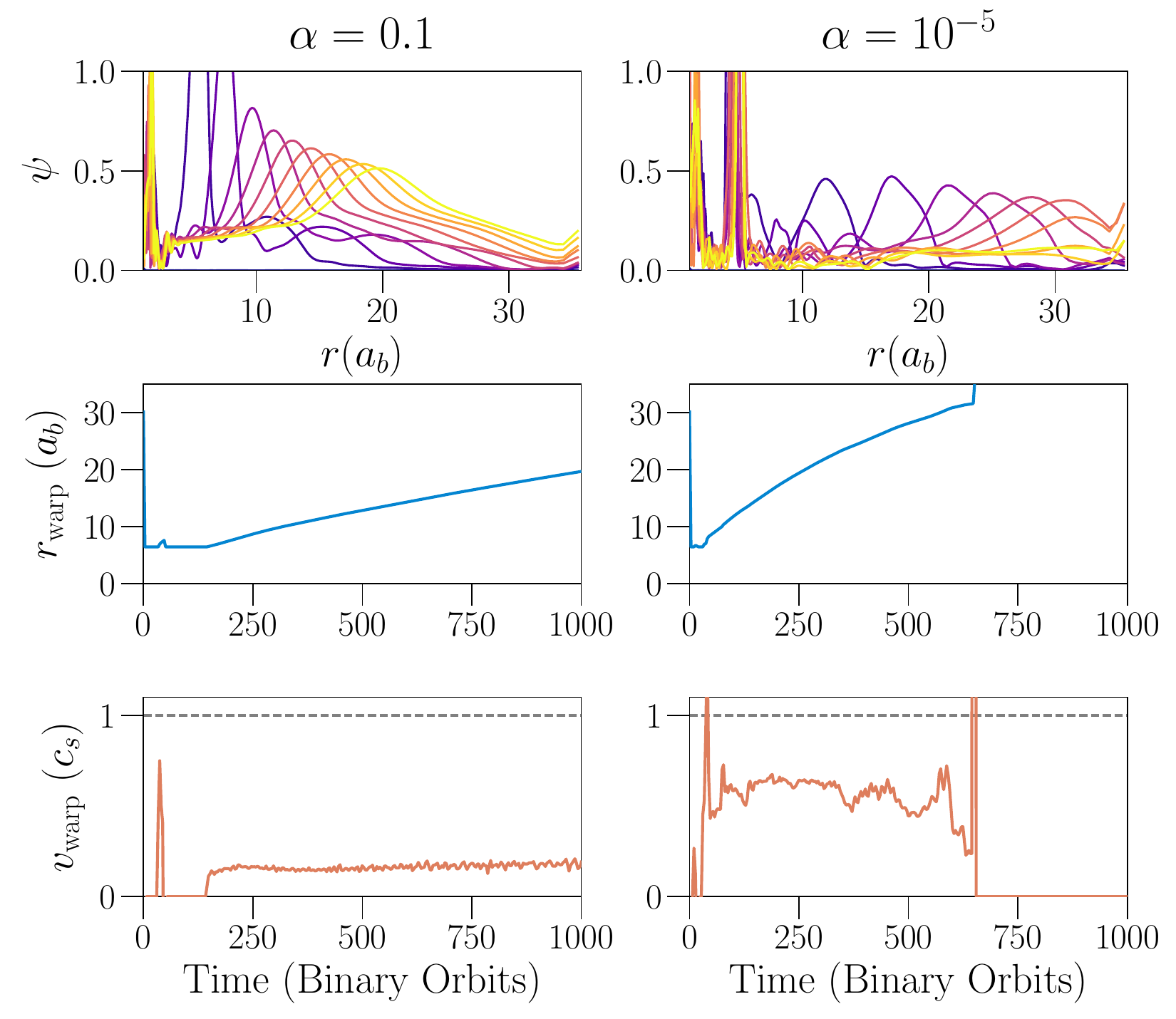}
    \caption{Motion of the warp for the viscous ($\alpha = 0.1$, left) and inviscid cases ($\alpha = 10^{-5}$, right).  \emph{Top row:} The $\psi$ parameter plotted against the disc radius.  Curves are plotted every 100 binary orbits as in Figure \protect\ref{fig:warpevo}.  \emph{Middle row:} The radial location of the warp, determined as the local location of maximum $\psi$. For wave-like propagation, both the initial break and secondary wave propagation are plotted.  \emph{Bottom row:} The outward propagation velocity of the warp.  The dotted line represents the local sound speed velocity of $c_{\rm s}$ at the warp radius. }
    \label{fig:warpspd}
\end{figure}

\begin{figure}
    \centering
    \includegraphics[width=\columnwidth]{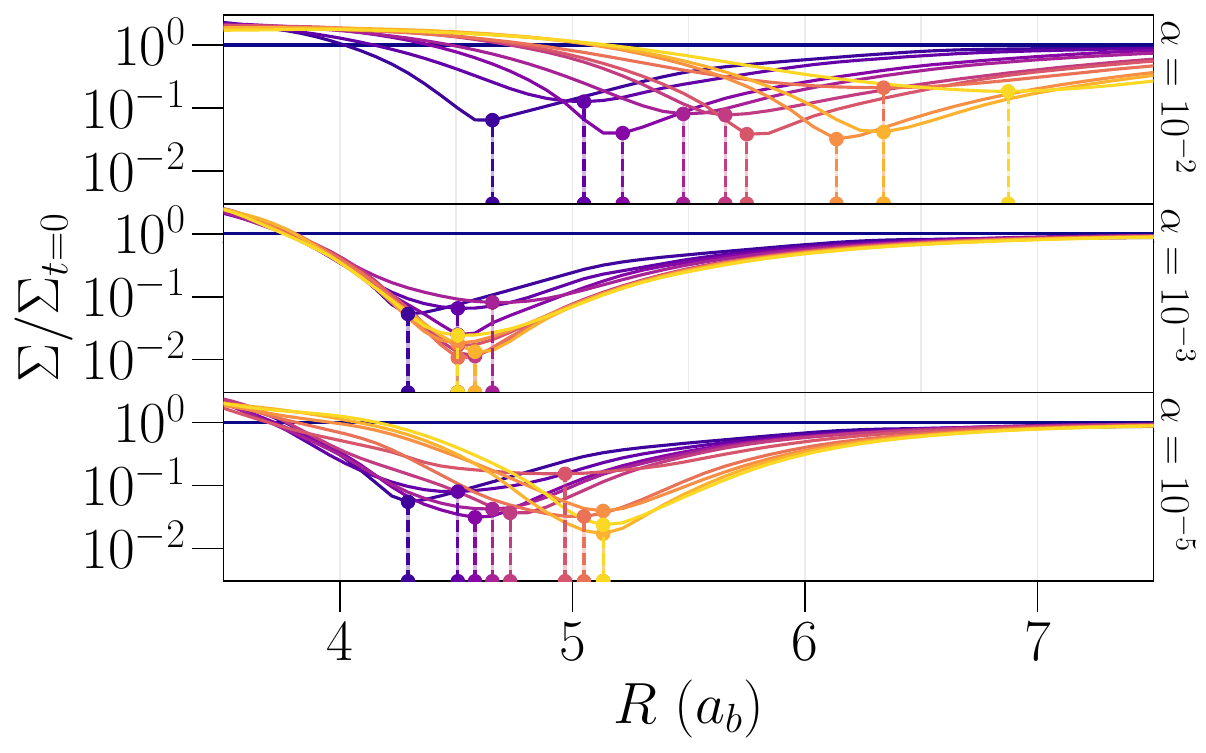}
    \caption{Drift of the breaking radius seen for our polar-aligning discs in the bending wave regime.  Each panel plots the normalized surface density of the disc every 100 binary orbits, with the same coloring as Figures \ref{fig:warpevo} and \ref{fig:warpspd}.  Vertical dashed lines indicate the location of the breaking radius for each snapshot.  }
    \label{fig:breakdrift}
\end{figure}

\begin{figure}
    \centering
    \includegraphics[width=\columnwidth]{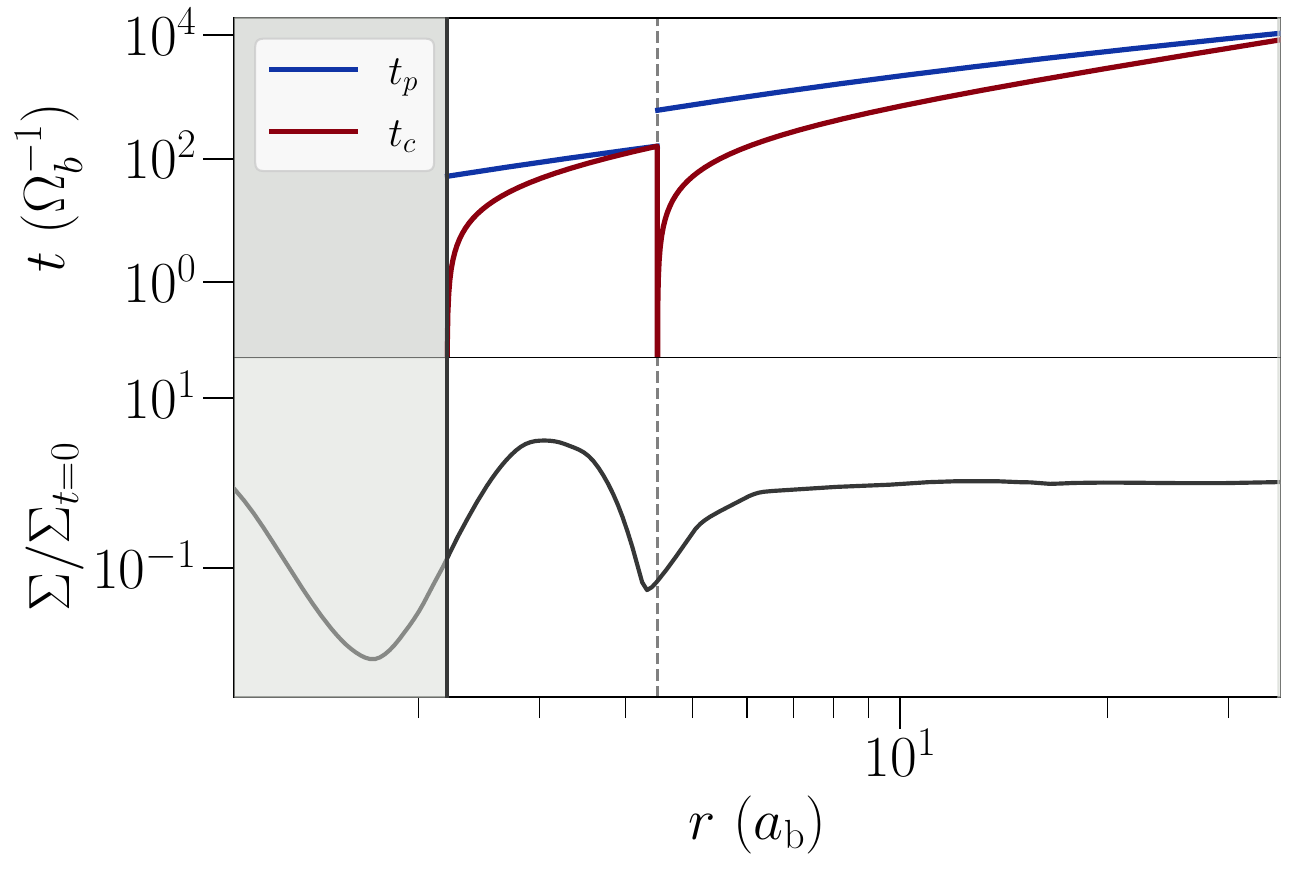}
    \caption{Comparison of the simulated breaking radius to analytical predictions. \emph{Top:} Analytical timescales as predicted from Equations (\ref{eq:tprec_nl}) and (\ref{eq:tcomm_nl}).  Here, $\rin$ is calculated to be $2.2a_{\rm b}$ (see text for details). \emph{Bottom:} Surface density of the $\alpha = 10^{-5}$ simulation at $t = 100$ binary orbits, in units of the initial surface density profile $\Sigma_0$.  The breaking radius is denoted by the sharp dip in $\Sigma/\Sigma_0$.  In both plots, vertical dashed lines indicate the predicted locations of the breaking radius.  }
    \label{fig:breakcomparison}
\end{figure}

Our polar-aligning disc simulations exhibit both diffusive and wave-like behavior, depending on the choice of $\alpha$.  Figure \ref{fig:polardensity} shows the various discs at $t = 1000$ binary orbits, displaying varied behaviors in their evolution.  For the $\alpha = 10^{-1}$ simulation, the disc develops a strong warp which evolves diffusively, while for the $\alpha = 10^{-2}, 10^{-3},$ and $10^{-5}$ simulations the disc breaks and exhibits wave-like dispersion.  Figure \ref{fig:warpevo} shows the time evolution for both regimes, following the evolution of the $z$-component of the unit angular momentum vector.  In all simulations, an initial warp is excited in the inner disc by the binary precession in the first $\sim 200$ binary orbits.  For the diffusive case, the inner regions of the disc quickly align towards polar due to the short alignment timescale, and the initial warp spreads outwards in the disc diffusively.  For the wave-like cases, the initial warp develops into a sharp discontinuity, characteristic of a disc breaking event.  After the disc breaks, a smaller disturbance propagates through the outer disc in a wave-like manner.  

%Examining the wave-like regime in Figure \ref{fig:warpevo} shows that the wavefront widens over time, indicating some amount of diffusivity is still present.  The degree of diffusivity is 

%Afterwards, the disk precession slows down (see Figure \ref{fig:angle}) and the warp spreads outwards diffusively.  The $\alpha = 10^{-3}$ warp simulation exhibits a different behavior.  After the initial warp excitation, the angular momentum shows a sharp discontinuity at roughly $R = 4 a_{bin}$, corresponding to a disk breaking event.  A separate warp structure propagates past the breaking radius in a mostly wave-like manner over time.  In this simulation, the inner disk separates from the outer disk and precesses independently the binary torques.

To determine the location for the warp and follow its movement within the disc, we track the position of maximum $\psi$ as the disc evolves.  In the case of a disc that breaks, two large peaks in $\psi$ are present, one at the breaking radius and another at the bending wave in the outer disc. We identify the radial location of these features by searching for the largest two maxima in $\psi$ to determine the locations of the disc breaking radius and the warp in the outer disc.  The warp location and propagation speed for our warp simulations is shown in Figure~\ref{fig:warpspd}.

For simulations in the diffusive regime, the inner disc generates a large warp as the disc angular momentum aligns to the precession axis.  The warp creates a large peak in $\psi$ that travels outwards at roughly 0.2 times the local sound speed.  The left-hand panels of Figures \ref{fig:warpevo} and \ref{fig:warpspd} show that the warp starts as a very steep and local feature, but broadens over time, encompassing a range of approximately $10a_{\rm b}$ by the end of the simulation.  A small wave is launched ahead of the main warp (visible in Figure \ref{fig:warpspd} as a series of secondary purple peaks between 10 to 25 $a_{\rm b}$, beneath the yellow curves drawn at later times).  This wave does not appear with a similar feature in Figure \ref{fig:warpevo}.  The highly diffusive nature of this disc quickly removes this feature, causing it to disappear by $t = 500T_{\rm b}$.

Simulations in the wave-like regime show the initial disc breaking at $r_{\rm break} \sim 4 a_{\rm b}$, independent of the small disc viscosity.  After the initial breaking event, the breaking radius moves outwards, faster for the intermediate case $\alpha = 0.01$.  The bottom panel of Figure \ref{fig:warpspd} shows that the outward warp decreases in velocity as it moves outwards in the disc, the propagation speed following close to one-half of the local sound speed in the disc.

We examine the drift of the breaking radius in Figure \ref{fig:breakdrift}.  Each panel shows the location of the breaking radius for a simulation with a particular $\alpha$ viscosity, with curves plotting the normalized surface density and vertical dashed lines marking the location of the breaking radius at each time.  Outward drift of the break is seen for the $\alpha = 10^{-2}$ and $10^{-5}$ simulations, but not for the $\alpha=10^{-3}$ simulation.  For the $\alpha=10^{-5}$ simulation, the break forms at roughly $r_{\rm break} = 4.25 \abin$ and slowly drifts outwards until $t = 800 T_{\rm b}$, where the break appears to stabilize at a final location of $5.1\abin$.  The $\alpha = 10^{-2}$ simulation breaks the disc slightly farther out, at $r_{\rm break} \sim 4.65 \abin$, and drifts outwards at a rate nearly twice as fast as the $\alpha = 10^{-5}$ simulation.  By the end of the simulation, the break has nearly reached $7\abin$, and does not show any signs of slowing down.  In both simulations, the vertical lines representing the location of breaking radius are unevenly spaced, suggesting that the breaking radius may not be growing constantly.  Large growth events may be related to the times when the inner and outer discs are momentarily aligned, allowing for increased accretion between the two discs.

\begin{figure}
    \centering
    \includegraphics[width=\columnwidth]{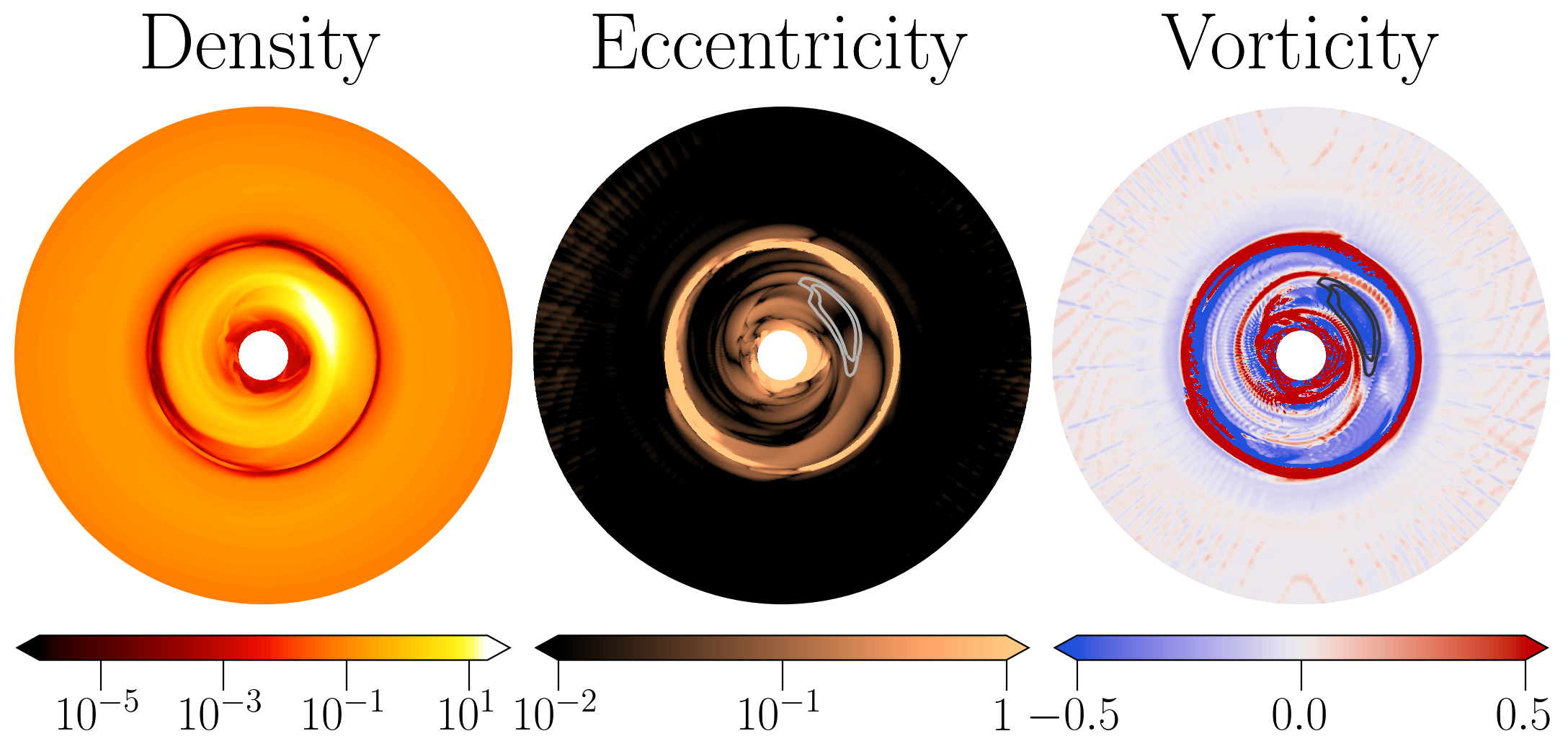}
    \caption{Inner regions of the $\alpha = 10^{-5}$ polar-aligning simulation, projected back onto a single plane.  \emph{Left:} Midplane density.  The inner disc features an RWI vortex.  \emph{Center:} Disc eccentricity.  The inner disc is somewhat eccentric.  \emph{Right:} Disc vorticity.  Note the strong vorticity minimum at the location of the vortex. }
    \label{fig:warpvorticity}
\end{figure}

Our analytical equations for disc breaking show close agreement with the location of the break observed in the numerical simulations.  In Figure \ref{fig:breakcomparison} we compare the location of the break in our $\alpha = 10^{-5}$ simulation to the location predicted by Equations~(\ref{eq:tprec_nl}) and~(\ref{eq:tcomm_nl}).  The breaking radius is visible as a sharp drop in the disc surface density.  We show the disc surface density at $t = 100$ binary orbits, just after the initial break has formed.  Comparing the location of the break at early times is important for accurately determining the location of the breaking radius, since the break slowly drifts outwards in all of our simulations.

To determine the value of $\rin$ for our analytic equations, we compare the density-weighted angular momentum of the disc $\int \Sigma r^3 \Omega dr$ to the analytic value expected for a disc with the same surface density power law profile.  For this calculation, we choose a snapshot shortly after the initalization, when the disc has had time to equalize but before a break has had time to open up.  We consider $\rin$ to be the location at which these two values diverge.  For the polar-aligning discs, this gives a value of $\rin = 2.2a_{\rm b}$.

We also observe substructures in the individual discs created after the breaking event.  Figure \ref{fig:warpvorticity} shows the $\alpha = 10^{-5}$ simulation as it would appear ``flattened'' onto a single plane, as well as the disc eccentricity and vorticity. Simulations with $\alpha = 10^{-3}$ and $\alpha = 10^{-2}$ show similar details in the gap regions, but no substructures in the inner disc.  We calculate the disc eccentricity and vorticity in a similar manner to that of \cite{Rabago2023}.  With low disc viscosities, the inner disc develops a large Rossby Wave Instability (RWI) vortex \citep{Lovelace1999,Li2000}, as well as a localized overdensity and a single-armed spiral.  This feature is identical to the vortices seen in the polar discs of \cite{Rabago2023}.  We observe less features in the outer disc; some faint two-armed spirals are visible at certain times, and no vortices are present.  The outer disc has undergone less dynamical times, and so may not have had enough time to develop a large vortex at its inner edge.  The gap edge between the inner and outer disc may not be as steep as the edge of the inner disc truncated by the central binary, which may also inhibit the outer disc from forming a strong vortensity minimum and the growth of RWI vortices \citep{Bae2015}.

%-----------------------------------------------%
%---------------- GW ORIONIS -------------------%
%-----------------------------------------------%
\begin{figure}
    \centering
    \includegraphics[width=\columnwidth]{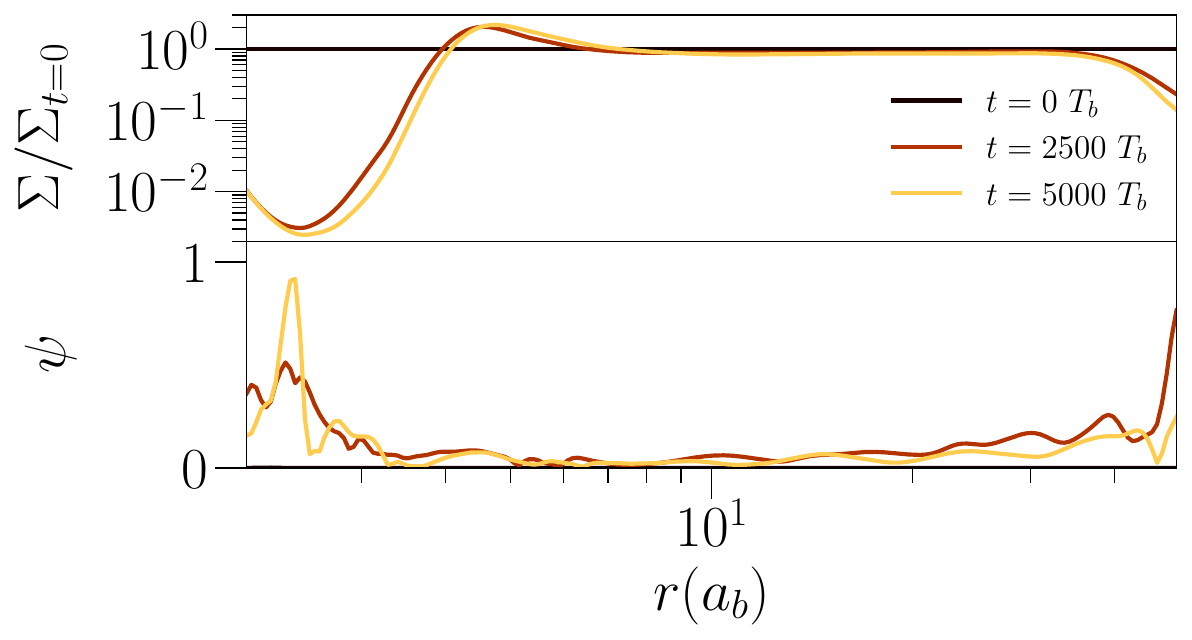}
    \caption{Surface density and warp profile of the GW Ori simulation at $t= 0, 2500$, and $5000 T_{\rm b}$.}
    \label{fig:sd_gwori}
\end{figure}

\subsection{GW Orionis}
\label{sec:results_gwori}
Figure \ref{fig:sd_gwori} shows the disc surface density and warp strength of the simulation representing the GW Orionis system (Sec.\ref{sec:methods_gwori}).  The disc develops a warp within the first 1000 binary orbits.  However, no disc breaking event occurs, even after the next several thousand orbits of the simulation.  A light warp persists in the disc throughout the simulation time, but is mostly confined to the inner regions of the disc, and never grows large enough to become unstable and evolve into a full break.  We run tests with different $\alpha$ viscosities, simulating out to at least 1000 binary orbits, but none of the simulations are able to create a break in the disc.

The arrangement of the GW Orionis binary changes the strength of the induced precession.  The terms in Equation (\ref{eq:precconst}) suggest that the precession induced by the binary is strongest for polar discs around high eccentricity, equal mass binaries.  The GW Orionis system has a moderate binary eccentricity of $e_{\rm b} = 0.379$ and a mass ratio, $q = M_C/M_{AB} = 0.348$, making the precession rate normalized by $\Omega_{\rm b}$ of this system about one-fifth as strong as that induced by the polar discs in Section \ref{sec:results_polar}.

In our simulations, the binary torque truncates the inner edge of the disc at roughly $\rin = 3.5a_{\rm b}$.  This is a larger value of $\rin$ than observed in the \cite{Smallwood2021} simulations, and larger than the value we initially used for calculating the breaking radius above.  The size of the inner cavity remains larger than the \cite{Smallwood2021} simulations, even for runs with the same $\alpha$-viscosity.  Recalculating the location of the breaking radius with the value of $\rin$ seen in the simulations changes the disc timescales so that no break occurs in the disc.

Although we try to reproduce the simulations of \cite{Smallwood2021} with a small inner cavity as best as possible, using nearly identical parameters for both the disc and binary, we are unable to reproduce the observed disc breaking using our grid-based simulations.  Our use of a slightly larger scale height due to computational limitations decreases the value of $t_{\rm c}$, but the analytic equations we present in Section \ref{sec:timescales} predict that we should still observe a break as long as the inner edge of the disc is truncated in the same location as the \cite{Smallwood2021} simulation.  As described in Section \ref{sec:timescales_gwori}, the initial surface density profile of the \cite{Smallwood2021} simulation is artificial; too much material is initialized close to the binary with a sharp cutoff, allowing the inner disc to precess and break before the inner cavity is fully cleared.  We were unable to replicate their initial density profile exactly in our grid-based simulations, as the inner cavity of our disk is a gradual truncation starting at $\rin = 24$au in order to ensure simulation stability.  This slightly increases the value of $t_{\rm p}$ and may be enough to prevent breaking.  It is also possible that differences in the code formulation may play a role in how the initial stages of the disc evolve, when the cavity depletes and the disc is most likely to break.

%\textbf{The Shakura-Sunyaev $\alpha$-viscosity varies within the SPH disc of \cite{Smallwood2021}, from $\alpha_{\rm SS} = 0.007$ to $0.018$ (see their Appendix).  Including contributions from the quadratic $\beta_{\rm SPH}$ term \citep{Meru2012} gives an $\alpha$-viscosity of $\alpha_{\rm SS, lin}+\alpha_{\rm SS, quad} = 0.011$ to $0.046$. A higher $\alpha_{\rm SS}$ value allows the disc to maintain a smaller inner cavity.}

%\SL{Are you trying to initially use the smaller Smallwood inner radius? But doesn't the inner radius grow by tidal clearing, so that no breaking occurs in the analytic model, as you say in the previous paragraph? } 

%\RGM{The simulation in Smallwood 2021 with a smaller inner radius is somewhat artificial. The simulation begins with a power law surface density profile with no smoothing at the inner edge, therefore the peak surface density is at r=20au. This is therefore not a realistic surface density profile. The disc breaks because too much material has been placed close to the binary.  I think this could explain why you cannot reproduce that simulation.  }

%-----------------------------------------------%
%--------------- MULTI BREAK -------------------%
%-----------------------------------------------%
\subsection{Multiple Breaks}
\begin{figure*}
    \centering
    \includegraphics[width=\textwidth]{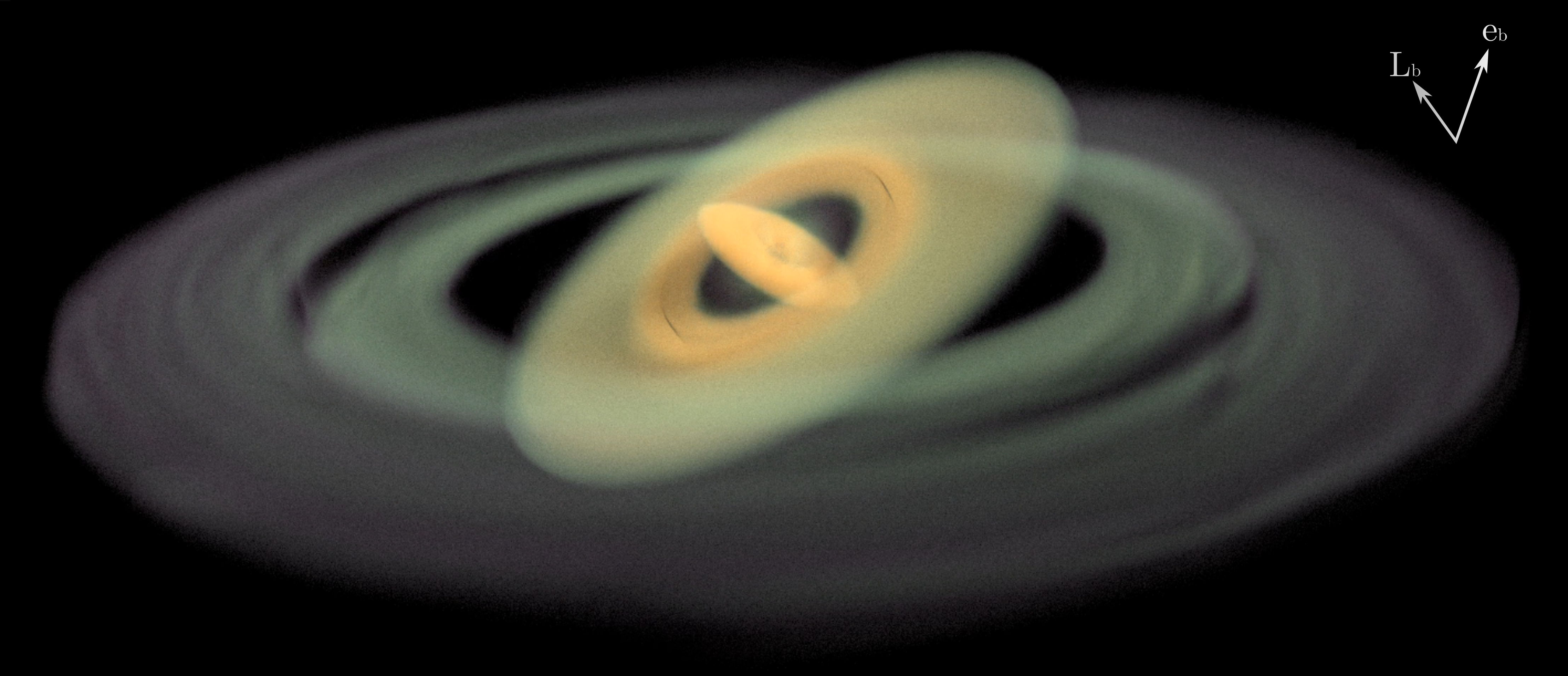}
    \caption{Rendering of the multiple break simulation at $t=1200$ binary orbits.  The directions of $\vecL_{\rm b}$ and ${\bm e}_{\rm b}$ are indicated in the top right, with $\vecL_{\rm b}$ pointing into the page.  The first two breaks at $7a_{\rm b}$ and $28 a_{\rm b}$ are visible.  There is a large warp at $35a_{\rm b}$ that appears similar to a break, but the disc is not broken at this distance.  See text for details.  An animation of the disc evolution is also available as an \href{https://youtu.be/_P7X3Wbf20Y}{online video}. }
    \label{fig:multibreak}
\end{figure*}

\begin{figure}
    \centering
    \includegraphics[width=\columnwidth]{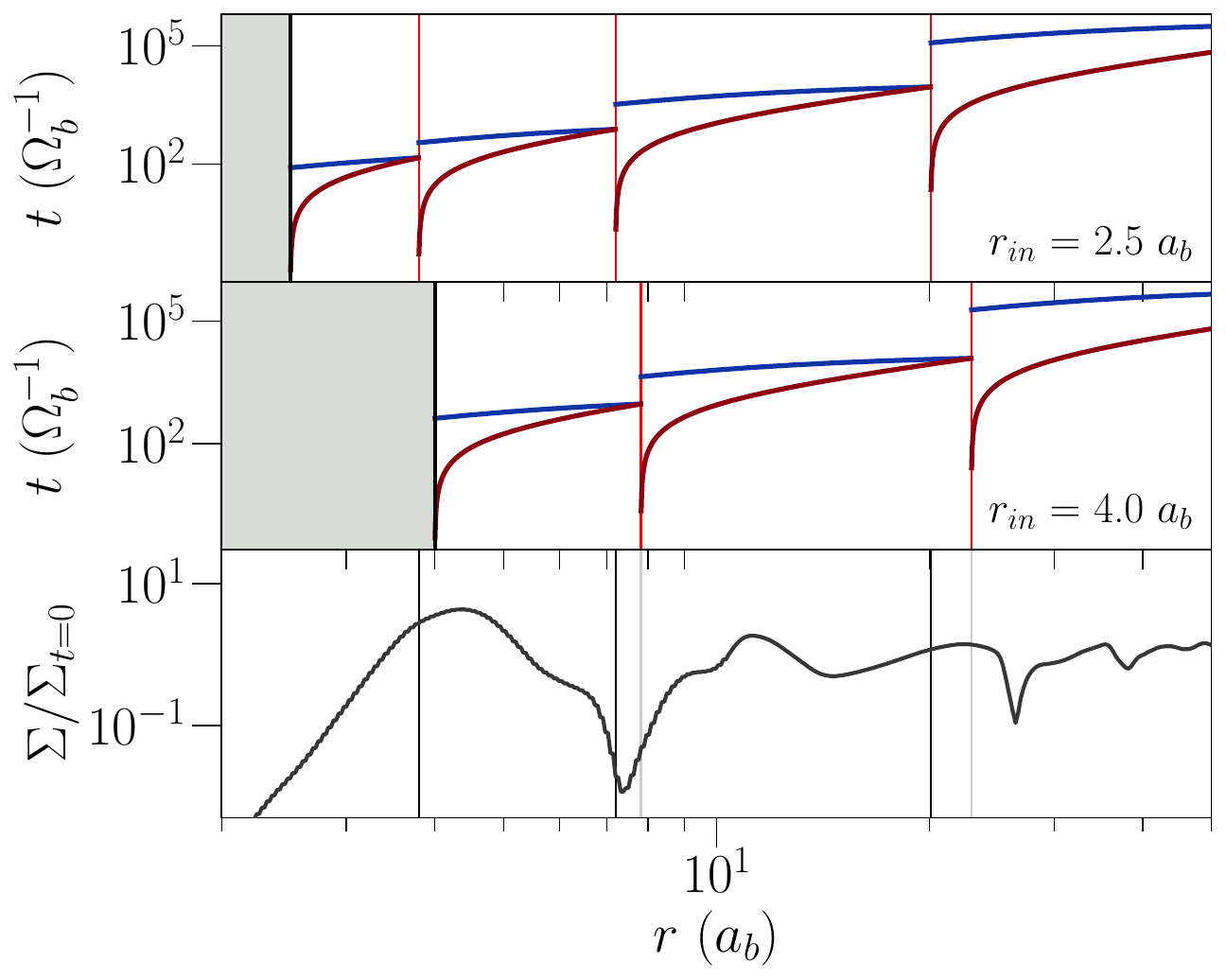}
    \caption{Similar to Figure \ref{fig:breakcomparison}, but for the multiple break simulation.  In this figure, each panel shows different choices of the inner radius, $\rin = 2.5 a_{\rm b}$ (top) and $\rin = 4.0 a_{\rm b}$ (middle).  Both choices produce outer breaking locations that are roughly consistent with the observed breaks in the simulation. The bottom panel is from a simulation at a time of 1500 $T_{\rm b}$. }
    \label{fig:breakcompmulti}
\end{figure}

\begin{figure}
    \centering
    \includegraphics[width=\columnwidth]{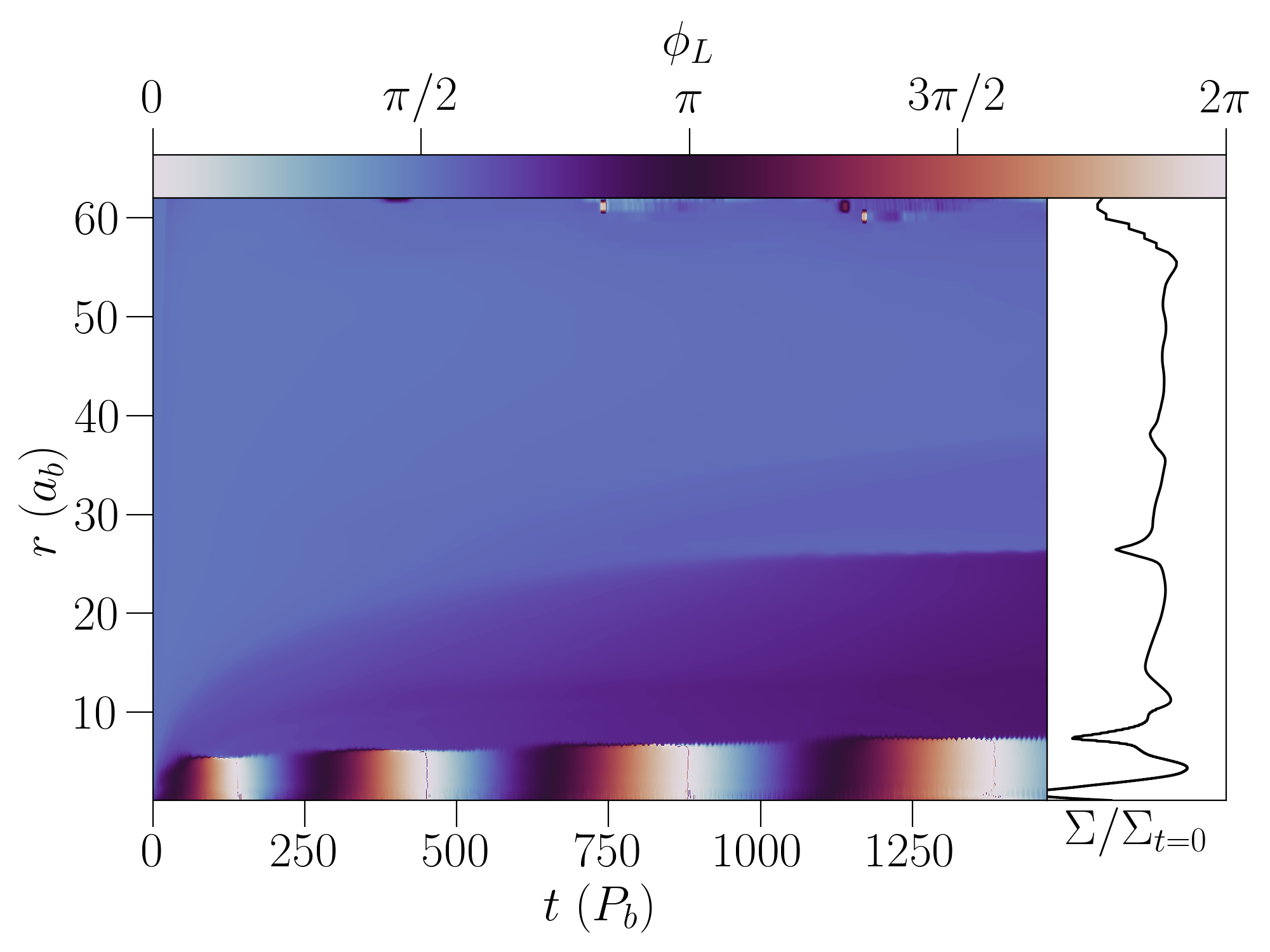}
    \caption{Azimuthal rotation angle plotted as a function of radial location and time.  Areas in the same connected disc precess together, and create a band of the same color.  By the end of the simulation, three distinct bands are visible, corresponding to the three discs in the simulation.  The disc surface density profile at $t=1500 T_b$ (same as  Figure \protect\ref{fig:breakcompmulti}) is also shown at right to compare the locations of the observed gaps.  A detailed time evolution of the disc surface density and warp profile is also available as an \href{https://youtu.be/GHrsw9brblc}{online video}. }
    \label{fig:multibreakangle}
\end{figure}

\label{sec:results_multi}
A visualization of our multiple break simulation (Sec. \ref{sec:methods_multi}) is shown in Figure \ref{fig:multibreak}.  Two disc breaking events are visible, separating the disc into three distinct rings.  At the end of the simulation, the disc breaks are located at radial distances of $7$ and $28 a_{\rm b}$, with the outer edge of the disc reaching equilibrium at roughly $\rout = 60a_{\rm b}$.  The third ring shows a strong warp at roughly $35a_{\rm b}$.  This warp is strong enough for the disc to appear broken due to how thin the disc is at this radius, but subsequent analysis will show that the disc is still communicating across this region.  Similar to the polar discs in Section \ref{sec:results_polar}, we see RWI vortices form in the innermost disc, but not in the second or third discs.

%Figure \ref{fig:breakcompmulti} compares the observed locations of each break to the locations predicted by the timescale equations.  Using the same method described in Section \ref{sec:results_polar}, we identify a similar inner boundary of $\rin = 2.7 a_{\rm b}$.   Our analytical equations predict 3 breaks, with the second and third breaks in rough agreement with the locations of the breaks observed in the simulation.  The first break, expected at $R = 4.25 a_{\rm b}$, is not observed in the innermost disk.  

Figure \ref{fig:breakcompmulti} compares the observed locations of each break to the locations predicted by the timescale equations.  We find it is difficult to determine a suitable inner radius for this disc.  Choosing an inner boundary of $\rin = 2.5 a_{\rm b}$ our analytical equations predict 3 breaks, with the second and third breaks in rough agreement with the locations of the breaks observed in the simulation.  The first break, expected at $r = 4.25 a_{\rm b}$, is not observed in the innermost disc.  Instead the observed breaks may be roughly fit by choosing $\rin = 4.0 a_{\rm b}$, but this places the ``inner edge'' of the disc much farther out, in a region where the local surface density is higher than the initial condition.  

To better identify the regions in the disc that are communicating, we calculate the azimuthal angle of the shell-averaged angular momentum vector, defined as
\begin{equation}
    \phi_L = \arctan{\left( \frac{L_y}{L_x}\right) },
\end{equation}
In our multiple break simulation, the $z$ direction is along the binary eccentricity vector and the binary orbit is in the $x-z$ plane, so $\phi_L$ can be viewed as the longitude of ascending node of the disc in the $x-y$ plane as it circulates around the binary eccentricity vector.
Regions of the disc in radial communication will precess together, and thus have roughly the same value of $\phi_L$.  In Figure \ref{fig:multibreakangle}, we plot the value of $\phi_L$ across the disc as a function of simulation time.  As the disc evolves and breaks off rings, distinct horizontal bands of color appear, each band corresponding to a separate disc in the final simulation.  Disc breaking events are visible as sharp boundaries between two horizontal bands, as the value of $\vecL$ changes sharply from one disc to another.

In Figure \ref{fig:multibreakangle}, it can be seen that the innermost disc breaks off almost immediately, in about 100 binary orbits.  The fast precession of the inner disc causes $\phi_L$ to cycle several times during the rest of the simulation, and creates the colored band at the bottom of the Figure.  The outer disc then develops a warp that gradually extends outwards to about $25 a_{\rm b}$, before breaking off a second disc at roughly $1000 T_{\rm b}$.  After the disc breaks at $25 a_{\rm b}$, the warp between the first break and the second break gradually vanishes and this disc region becomes co-planar. This is another indication that the disc loses the communication at the breaking radius and the outer disc cannot torque the inner disc any more.  A faint band can be seen extending from the second disc to about $35 a_{\rm b}$ in the outermost disc, signifying a strong warp.  However, since a sharp discontinuity has not yet formed, this band is not yet considered a third breaking event.

The multiple break simulation showcases the difficulty in using Equations (\ref{eq:tprec_nl}) and (\ref{eq:tcomm_nl}) to predict the exact location of the breaking radii when multiple breaks are packed together in a small radial range.  Breaking events require a few local dynamical timescales to fully develop, so the innermost break will have time to drift outwards while subsequent breaks form and develop.  The bottom panel of Figure \ref{fig:breakcompmulti} shows that the surface density, normalized to an initial power-law distribution, is quite disturbed even in regions far away from the breaks, suggesting that the disc no longer follows a power-law distribution even though the analytic equations assume one when calculating the location of a break.  We discuss this behavior further in Section \ref{sec:discussion_equations}. 

%%%%%%%%%%%%%%%%%%%%%%%%%%%%%%%%%%%%%%%%%%%%%%%%%%

%%%%%%%%%%%%%%%%%%% DISCUSSION %%%%%%%%%%%%%%%%%%%

%%%%%%%%%%%%%%%%%%%%%%%%%%%%%%%%%%%%%%%%%%%%%%%%%%
\section{Discussion}
\label{sec:discussion}

\begin{figure}
    \centering
    \includegraphics[width=\columnwidth]{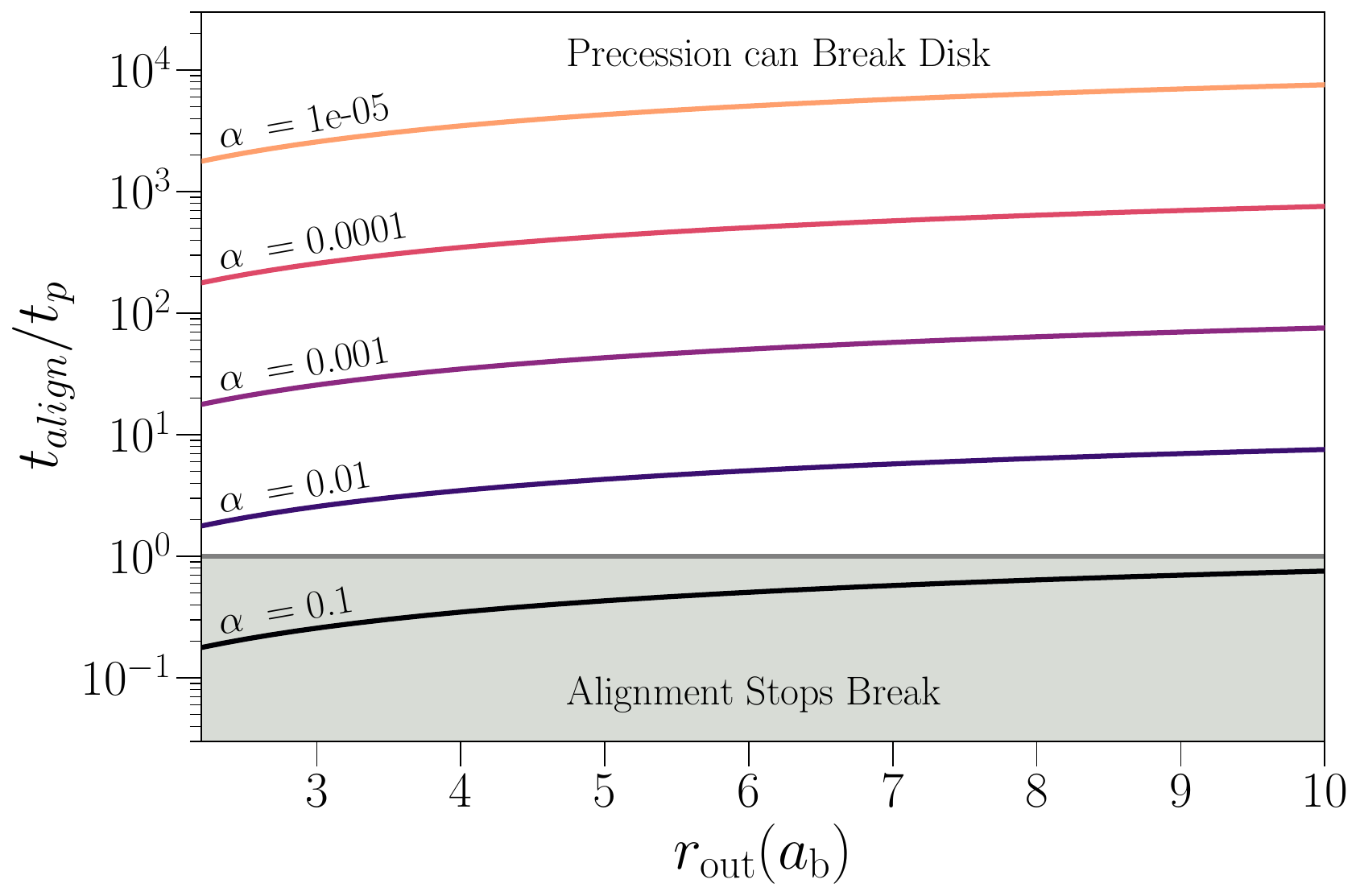}
    \caption{  $\mathbf{ \tau = t_{\rm align}/t_{\rm p} }$ as calculated by Equation (\ref{eq:tau}) for different values of $\alpha$.  The vertical line indicates the rough location of the break as seen in our low viscosity simulations. }
    \label{fig:tau}
\end{figure}

For the polar discs used in Section \ref{sec:methods_polar}, we show the effects of disc viscosity by plotting $\mathbf{ \tau = t_{\rm align}/t_{\rm p} }$ as a function of radius in Figure \ref{fig:tau}, with the assumption that while calculating $t_{\rm p}$ all material interior to $r$ is able to precess coherently.  We see that for our selected disc parameters, the discs with $\alpha \lesssim 0.01$ maintain $\tau > 1$ for the entirety of the disc and thus are able to break under the induced precession of the binary.

We can estimate $\nu_2$ and $\nu_3$ for our $\alpha = 0.1$ simulation using Figures 4 and 5 of \cite{Ogilvie1999}.  From our Figure \ref{fig:warpspd}, the maximum warp amplitude is roughly $\psi_{\rm max} = 0.5-0.75$, at a distance of $r = 10 a_{\rm b}$.  This corresponds to warp viscosities of $\nu_2 = 0.019 a_{\rm_b}^2\Omega_{\rm b}$, $\nu_3 = 0.0013 a_{\rm_b}^2\Omega_{\rm b}$, and $t_{\nu_2} = 5000 \Omega_{\rm b}^{-1}$.

We note that, for the simulations in Section \ref{sec:results_polar}, there is a change in the overall disc orientation of about 5 to 10 degrees from its initial orientation.  This change occurs faster than the local precession timescale induced by the binary, and instead occurs as the warp (diffusive or wave-like) propagates through the disc.  This effect is partially visible in Figure \ref{fig:polardensity}, since the disc, initialized in alignment with the simulation $x$-axis, is now misaligned at the end of the simulation.  We consider this to be an effect of the $\nu_3$ warp viscosity, which has caused the entire disc to precess.  

%Despite observing these effects, calculating the values of $\nu_3$ (as well as the warp coefficient $\nu_2$) is not straightforward; previous works such as \cite{Lodato2007} have required a separate solving and fitting of the differential equations to determine them, which is beyond the scope of our work here.

%Using an outer radius of $r = 5a$, a typical value for the disk precession time from Equation \ref{eq:tprec_nl} is on the order of hundreds of binary orbits.  For the disks used in Section \ref{sec:methods_polar}, this gives a value of $\tau \sim 1/2\pi\alpha$

\begin{figure}
    \centering
    \includegraphics[width=\columnwidth]{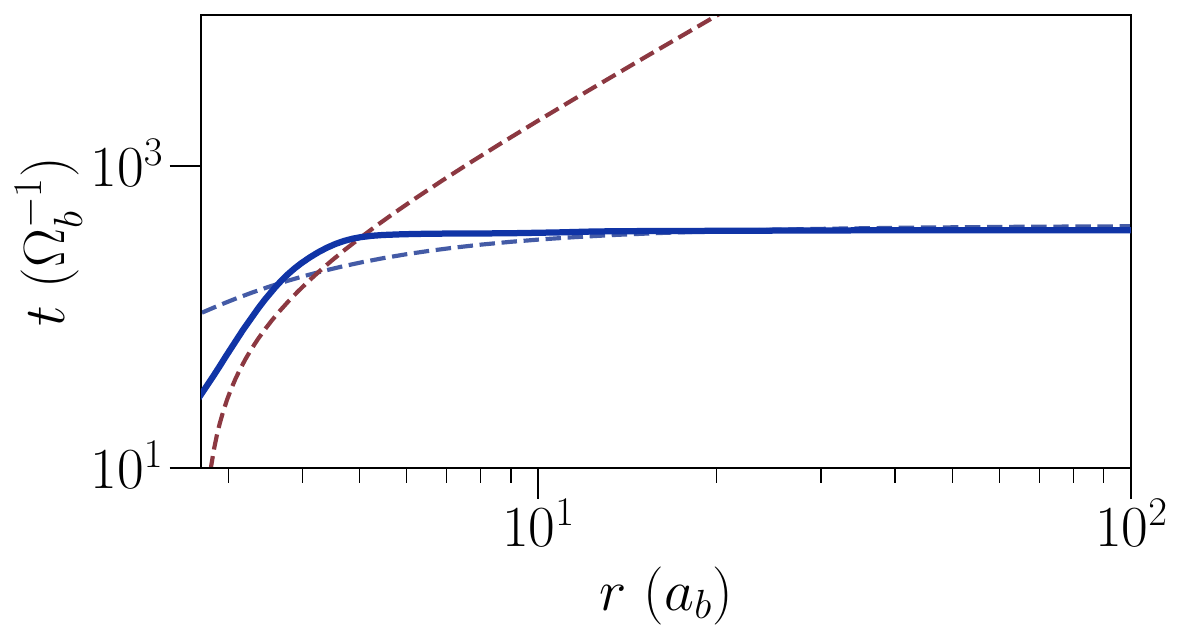}
    \caption{Comparison of disc timescales using the analytic equations Eq. (\protect\ref{eq:tprec_nl}) and (\protect\ref{eq:tcomm_nl}) to the precession time calculated by directly integrating Eq. (\protect\ref{eq:diskfreq}) with the observed surface density. Dashed red and blue lines indicate the analytic solutions for $t_c$ and $t_p$, while the solid blue line shows the directly integrated value for $t_p$.}
    \label{fig:breakcomp_integration}
\end{figure}

\subsection{Efficacy of the Disc Breaking Equations}
\label{sec:discussion_equations}

The location and number of breaking radii determined by Equations~(\ref{eq:tprec_nl}) and~(\ref{eq:tcomm_nl}) is a sensitive function of the disc parameters, particularly the inner disc radius $\rin$.  Because of this sensitivity, small adjustments in the value of $\rin$ result in large changes in the breaking radius.  Conversely, small adjustments in the location of the breaking radii only require minor changes in the required value of $\rin$.  An important caveat of this behavior is that it easily allows our description of disc breaking to remain consistent with observed discs, but it is harder for the same equations to act as predictive tools.  For example, Equations~(\ref{eq:tprec_nl}) and~(\ref{eq:tcomm_nl}) can be made consistent with simulations or observations by finely adjusting the disc parameters to match the observed location(s) of the breaking radius, but it is more difficult to predict the location of emergent breaks starting from the initial disc conditions.

%Our observed breaking radii in Section \ref{sec:results} confirm this behavior.  For each of our simulations, the expected breaking radii are usually significantly different as the inner radius shifts from our initial guess.  Once we adjust the value of $\rin$ to account for this change, the predicted breaks are much closer.

When examining our polar disc simulations in Section \ref{sec:results_polar}, we initially find the expected breaking radii are different from our initial guess as the inner radius shifts and reaches a quasi-steady state.  Once we adjust the value of $\rin$ to account for this change, the predicted breaks are in much closer agreement.  For the multiple break simulation (Section \ref{sec:results_multi}), we have a harder time trying to match both the observed breaking locations and the location of the inner radius simultaneously, especially when multiple breaks are packed in a small radial range.
%a given set of disk paramaters can be finely adjusted to give values of the breaking radius consistent with the locations seen in observations or simulations without causing a disagreement in the value of $\rin$.  

%Given a set of disk parameters, either from simulations or observations, these equations generate a value for the breaking radius that is generally close to the observed location.  A small adjustments to the input values, consistent with observational errors or differing definitions for radial cutoff, can be carefully adjusted to produce a breaking radius that is consistent with the observed breaking location.  However, this same sensitivity, primarily in that of the inner disk radius $\rin$, makes the reverse process is much more difficult.

The nature of this sensitivity leads to some difficulty in using Equations~(\ref{eq:tprec_nl}) and~(\ref{eq:tcomm_nl}) to predict the location of disc breaks.  There is no clear location at which the gas at the inner radius of the disc acts as an ``edge'' and stops precessing as a solid body with the rest of the disc.  Thus, different methods of calculating the inner radius such as using surface density cutoffs, precession rates, angular momentum deviations, or other methods will result in slightly differing values of $\rin$, leading to greatly differing estimates of the breaking radii in the disc.  Differences in $\rin$ caused by code formulation, such as those present in the GW Orionis simulations in Section \ref{sec:results_gwori}, can further complicate this issue.  This behavior was also noted in the simulations by \cite{Young2023}, who comment that the wide dependence on disc and binary parameters make the creation of a disc breaking equation a challenging matter.
%Thus, different definitions of $\rin$ using surface density cutoffs, precession rates, or other methods will calculate slightly different values, which will lead to greatly differing estimates of the breaking radii.  

%Our simulations of GW Ori in Section \ref{sec:results_gwori} show that the differences in code formulation (SPH vs. grid-based) can also change the location of $\rin$, further complicating this issue.  This behavior was also noted in the simulations by \cite{Young2023}, who comment that the wide dependence on disc and binary parameters make the creation of a disk breaking equation a challenging matter.

Our analytic equations are also restricted by their lack of full coverage of the possible parameters.  Previous SPH simulations have observed a dependence of disc breaking with inclination to the precession vector \citep{Facchini2013,Nealon2015}, suggesting that disc breaks are more likely for highly inclined or retrograde discs, and may not occur if the mutual inclination between the disc and precession vectors (prograde or retrograde) is small.  Analysis by \cite{Dogan2018} and \cite{Raj2021} suggest that a warp may continue to grow unstably if the warp amplitude $\psi$ is beyond a critical limit $\psi_c$, dependent on the disc viscosity.  The current analysis is relatively simple and is strictly analytic, using the initial power-law profiles set at the beginning of the simulation to calculate the breaking radius.  As the binary-disc system evolves, the gas may settle towards different power-law profiles than those set by the initial conditions, or towards a radial distribution not approximated by a power-law profile, as seen in Figure \ref{fig:breakcompmulti}.  An example of this is shown in Figure \ref{fig:breakcomp_integration}, where we compare the analytic, power-law disc timescales in our multiple break simulation to the timescales calculated by directly integrating Equation \ref{eq:diskfreq} (since our simulations are nearly isothermal, direct integration for $t_c$ is identical to the analytic result of Eq. (\ref{eq:tcomm_nl})).  There is a slight outward shift in the location of the breaking radius in this case, but the new location is similar to the location predicted by the analytic equations.  However, this adjustment may lead to cascading outward shifts for the subsequent breaks farther out in the disc.

\begin{figure}
    \centering
    \includegraphics[width=\columnwidth]{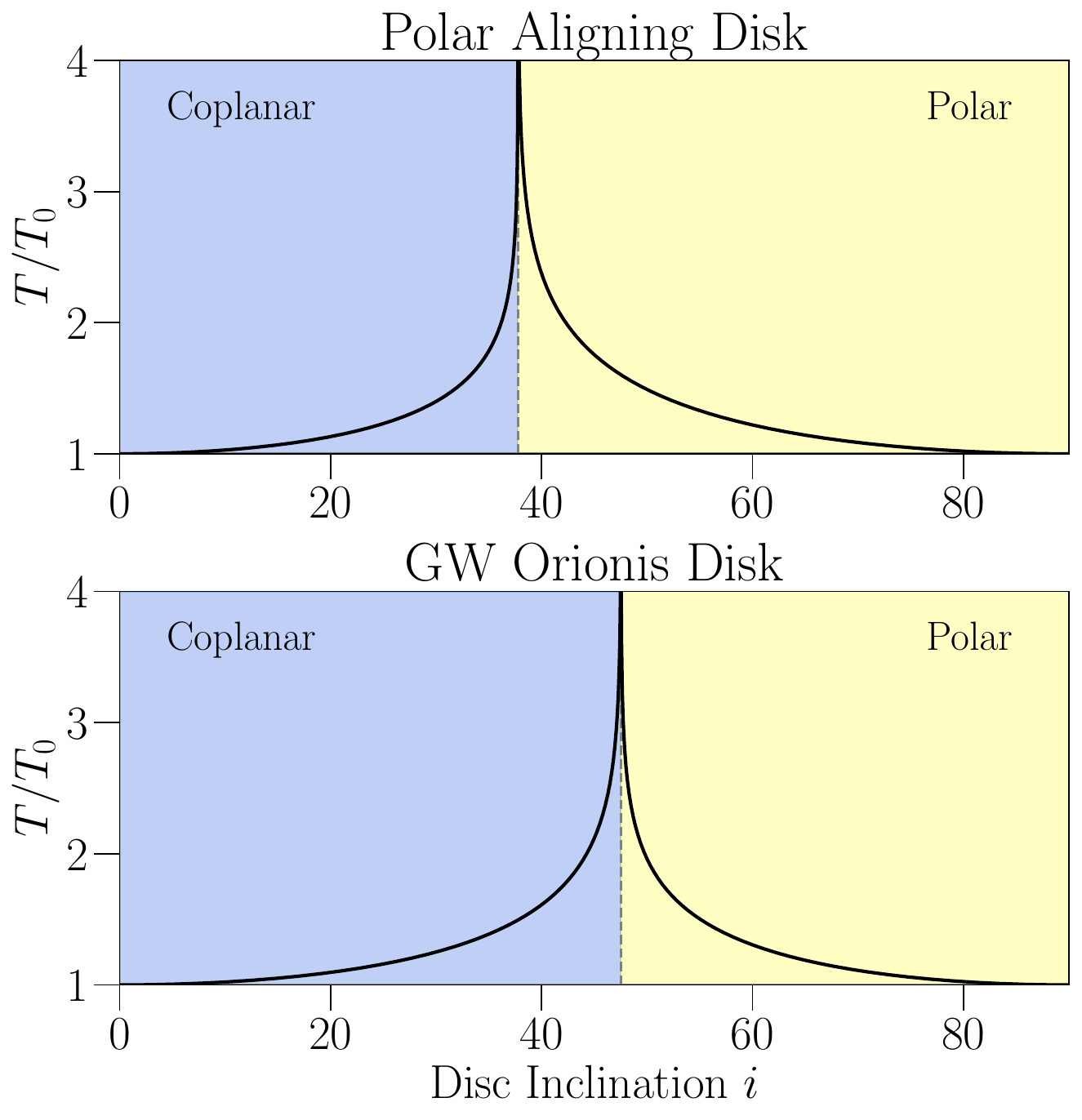}
    \caption{Analytic precession rates calculated from Eq. (2.32) of \protect\cite{Farago2010} as a function of disc inclination, for the cases of the polar-aligning discs in Sec.\ref{sec:results_polar} (top) and the GW Ori system in Sec.\ref{sec:results_gwori} (bottom).  The vertical line denotes the critical inclination $i_{\rm crit}$ for this system; inclinations lower than $i_{\rm crit}$ align to a coplanar orientation, while inclinations higher than $i_{\rm crit}$ align to a polar orientation.  Precession rates are normalized to $T_0$, the precession rate for a nearly coplanar or nearly polar disc.}
    \label{fig:analyticprecession}
\end{figure}

We can estimate the effects of the disc inclination on our results by examining the analytic equations of \cite{Farago2010} describing the orbits of test particles around eccentric binaries. We determine the correction factor for the particle precession period due to inclination away from nearly polar or coplanar orbits. This same correction factor also applies to a flat disc, since the factor is independent of radius.
For a circumbinary test particle, the precession period of the particle is given in Equation (2.32) of \cite{Farago2010} as
%\begin{equation}
%    T = \frac{16}{3n_1} \frac{M_{01}}{\beta_1} \left(\frac{a_2}{a_1} \right)^{7/2} \frac{K(k^2)(1-e_2^2)^2}{\sqrt{(1-e_1^2)(h+4e_1^2)}},
%    \label{eq:FL10precession}
%\end{equation}
\begin{equation}
    T = \frac{16}{3n_1} \frac{M_{\rm tot}}{\beta} \left(\frac{a_{\rm p}}{\abin} \right)^{7/2} \frac{K(k^2)(1-e_{\rm p}^2)^2}{\sqrt{(1-\ebin^2)(h+4\ebin^2)}},
    \label{eq:FL10precession}
\end{equation}
%\textbf{where $n_1$, $M_{01}$, and $\beta_1$ are the mean motion, total mass, and reduced mass of the binary ($n_1 = 1$, $M_{01} = M_{\rm tot}$ in our paper) \RGM{I think you should just use your symbols in equation 17}, and the orbital elements $a_1, e_1$ and $a_2, e_2$ refer to the semi-major axis and eccentricity of the binary and particle, respectively.  The constant $h = -4e_1^2 + \sin^2{i}(1+4e_1^2)$ describes the orientation of the test particle's orbit, and $K(k^2)$ is an elliptic integral of the first kind integrated over the azimuthal angle $\phi$, defined as  }
where $n_1$ is the mean motion of the binary ($n_1 = 1$ in our setup), and the orbital elements $a_{\rm p}$ and $e_{\rm p}$ refer to the semi-major axis and eccentricity of the orbiting particle.  The constant $h = -4\ebin^2 + \sin^2{i}(1+4\ebin^2)$ describes the orientation of the test particle's orbit, and $K(k^2)$ is an elliptic integral of the first kind integrated over the azimuthal angle $\phi$, defined as
\begin{equation}
    K(k^2) = 
    \begin{cases}
        \int_0^{\pi/2} \frac{1}{1-k^2 \sin^2\phi} d\phi & k^2 < 1 \\
        \int_0^{\phi_0} \frac{1}{1-k^2 \sin^2\phi} d\phi & k^2 > 1, 
    \end{cases}
\end{equation}
%\textbf{where $k^2 = \frac{5\e^2}{1-e_1^2} \cot^2 i$ and $\phi_0 = \arcsin\left(1/k\right)$.  The dependence of $K$ and $h$ on the inclination angle causes the value of $T$ to vary with changing disc inclination.}
where $k^2 = \frac{5\ebin^2}{1-\ebin^2} \cot^2 i$ and $\phi_0 = \arcsin\left(1/k\right)$.  The dependence of $K$ and $h$ on the inclination angle causes the value of $T$ to vary with changing disc inclination.

The effects of the change in precession time are shown as a function of disc inclination in Figure \ref{fig:analyticprecession} for the polar-aligning setup in Section \ref{sec:results_polar} and the GW Orionis setup in Section \ref{sec:results_gwori}.  The precession time lengthens considerably only in the vicinity of the critical inclination, and so we consider the effect on our simulations to be relatively small. The precession period diverges logarithmically  with the difference in inclination from the critical inclination.  Importantly, Figure \ref{fig:analyticprecession} shows that for discs within a few degrees of the critical inclination the disc will precess very slowly and is therefore unlikely to break.  Simultaneously, inclinations close to 0 or $90^\circ$ are also unlikely to break, since they are aligned close to their precession vectors and will not produce a strong warp amplitude.  Thus, certain inclinations between $0^\circ$ and $i_{\rm crit}$ or $i_{\rm crit}$ and $90^\circ$ are the most likely orientations for disc breaking.

The dependence on inclination derived here is different than that expected from \cite{Nixon2013}, who predict that breaking is most efficient for inclination angles between $45^\circ < i < 135^\circ$ and is strongest at $45^\circ$ and $135^\circ$.  This criterion is made by considering the cancellation of angular momentum from counter-aligned rings produced as the inner disc precesses.  Although this range of inclination angles is correct when considering circular binaries and Lense-Thirring precession induced from black holes, the addition of librating orbits for eccentric central binaries produces new regions at potentially moderate inclinations where disc breaking is unlikely.

%Our current equations do not include any factors related to these parameters, and so do not capture any related effects when determining a breaking radius.  The current analysis is relatively simple and is strictly analytic, using the initial power-law profiles set at the beginning of the simulation to calculate the breaking radius.  As the binary-disc system evolves, the gas may settle towards different power-law profiles than those set by the initial conditions, or towards a radial distribution not approximated by a power-law profile, as seen in Figure \ref{fig:breakcompmulti}.  Future analysis may improve this method by directly integrating the disc radial profiles at later times to provide better calculations of the disc timescales.

\subsection{Observational Signatures}

Our simulations of GW Orionis in Section \ref{sec:results_gwori} were conducted using a binary-disk inclination of $38^\circ$, as listed in previous works \citep{Kraus2020,Bi2020,Smallwood2021}.  After the simulations were conducted, we reanalyzed the observational data collected in \cite{Kraus2020} and found incorrectly listed values for the mutual inclination.  We recalculate these values using the data listed in their Tables S5-S6 and find the mutual inclinations of the outer AB-C binary to the R1, R2, and R3 dust rings (largest to smallest) to be $\theta_{\rm R1}=28.8^\circ$, $\theta_{\rm R2}=28.2^\circ$, and $\theta_{\rm R3}=24.9^\circ$.  These values suggest a slightly different structure of the GW Orionis disc than has previously been reported in \cite{Bi2020}.  In this model, the inclination of the disc does not change much with respect to the AB-C binary orbital plane, but the longitude of the ascending node $\Omega$ changes significantly (measured in the plane of the AB-C binary) between the R2 and R3 rings.  The reduction of the assumed misalignment angle keeps the GW Ori disc in the regime for nodal circulation and coplanar alignment, so we expect the previous and current simulations of the GW Ori disc to remain broadly consistent. This disc model is consistent with a disc in a mildly noncoplanar orientation that has been warped by the precession around the outer binary, with possible breaking near the location of the R3 dust ring.  Our new disc model better supports the possibility of disc breaking from azimuthal twisting motions rather than a changing binary-disc inclination as one moves inwards from R1 to R3.  The same misalignment at the inner R3 ring may imply that the alignment timescale is long relative to the disc precession timescale; from Equation \ref{eq:tau}, this implies a value of $\tau < 1$ for GW Ori, allowing the disc to break without impedance from viscous forces.
%The possibility of disc breaking is also supported by the near-constant value of the binary-disc inclination as one moves inwards from R1 to R3;

%The possibilty of disc breaking is also supported by the near-constant value of the binary-disc inclination as one moves inwards from R1 to R3; the same misalignment at the inner R3 ring may imply that the alignment timescale is long relative to the disc precession timescale.  From Equation \ref{eq:tau}, this implies a value of $\tau < 1$ for the GW Orionis disc, allowing the disc to break without impedance from viscous forces.  }

The changing orientation of warped discs is known to produce specific observable signatures.  Warping changes the projection of the disc onto the sky plane, and produces twisted or S-shaped contours when viewed in molecular lines.  This feature has been observed in the GW Orionis circumtriple system \citep{Bi2020,Kraus2020} as well as within the cavities of some transition discs such as HD 142527 \citep{Casassus2015}, and can be reproduced using simulations of warped discs \citep{Juhasz2017,Smallwood2021}.  Channel maps can also explore details about the larger surrounding structures around a warped disc, and reveal if an observed warp is caused by a central source such as a binary or if the warp is related to infalling matter from the surrounding molecular cloud.  In the case of GW Orionis, the identification of at least one streamer connected to the disc indicates both effects may be at play \citep{Czekala2017,Fang2017}.  High resolution kinematics of warped discs may be used to examine radial flows between broken discs and determine the strength of a disc warp, including if an observed warp is also a break.

%High resolution kinematics of warped discs may be able to determine the strength of a disc warp, and may aid in determining if an observed warp is also a break, as well as the flows in the region between the individual discs.

Broken discs have similar observational signatures.  If the inner and outer discs are in different orientations, each disc will have differently oriented and possibly separated butterfly patterns in CO channel maps \citep{Zhu2019}.  The inner disc will cast shadows as a pair of dark lanes onto the outer disc.  The orientation of these lanes may not be symmetric around the disc, and their pattern speed around the disc may not follow the precession rate of the inner disc, depending on the relative orientations of the inner disc, outer disc, and binary \citep{Facchini2018,Zhu2019}.  Discs with multiple breaks may have a complex series of shadowing features, as each disc shadows all of the discs external to it.  However, \cite{Facchini2018} find that the flaring geometry of the disc is an important feature for producing asymmetric illumination of the outer disc when the inner and outer discs have a low relative inclination.  Since multiple disc breaking is less favored for flaring discs, the shadowing patterns from these discs may appear more symmetric.  The local temperature changes caused by these shadows can generate spiral arms in the outer disc \citep{Montesinos2016}, so the thermal effects of these lanes may be important when attempting to reproduce observations.

The long-term movement of these shadows can give insights to the orientation of the central disc relative to the binary.  From Equation~(\ref{eq:precconst}), discs in coplanar and polar orientations have opposite signs of $k$, and will precess in opposite directions compared to their precession axis.  Long-term monitoring of discs, combined with rotational information from kinematic observations, can be used to determine the direction of the shadow's movement, and thus the precession direction of the inner disc.  This can be used to determine whether the inner disc is in a coplanar or polar configuration, and can place constraints on the arrangement of the central binary.

Once a break is formed in the disc, it may slowly drift outwards over time.  From Figure \ref{fig:breakdrift}, the average drift rate is roughly $10^{-3} a_{\rm b}/T_{\rm b}$ for the inviscid simulations, and roughly twice as fast for the intermediate $\alpha = 10^{-2}$ simulation.  It is unclear if this behavior is part of the disc settling towards a new equilibrium state, or if it is part of the long-term evolution of the disc after it has broken.  In our bending-wave simulations (Sec.\ref{sec:results_polar}), we observe a range of behaviors, from breaks which drift outwards for thousands of binary orbits to breaks that do not drift at all, with a relationship that is not monotonically dependent on the disc viscosity.  

For the break observed in the GW Orionis system, the drift rates listed above translate to an outward movement of around $0.7 - 1.4 \times 10^{-3} \rm{au/yr}$.  If the break continues outward at a constant rate, it will reach the edge of the observed disc (500 au, \citealt{Kraus2020}) in roughly $10^5$ years.  This may be relatively short compared to the disc lifetime of 1-10 Myr \citep{Andrews2009}, and may imply that broken discs are transient features.  On the other hand, the precession of the binary is constantly warping the disc, and should constantly generate breaks in the disc as long as it is still misaligned, similar to how other disc substructures such as gaps and spirals may be continuously generated by companions or orbiting planets.  In this way, it may be possible for a disc to generate multiple breaks even though the disc conditions are only enough to satisfy a single break, as breaks move outward in the disc and new ones are generated at the breaking radius.

%Recently, the conditions of disk breaking have also been studied by \cite{Young2023}, who use a suite of SPH simulations to examine under which conditions a circumbinary disk is likely to break.  Their conclusions for the causes of disk breaking are closely matched with ours; a higher binary mass ratio, thinner disk scale height, steeper power-law slopes, eccentric binaries, and sensitivity of the inner radius.  They are also able to determine a dependence on disk inclination, where the inclination to the precession vector must be somewhat high, $i \gtrsim 40^\circ$ for a break to occur.

%Slight differences in the formulations of SPH and grid-based codes may change the location of $\rin$, where the disk is truncated by the binary torque.  Since Equations \ref{eq:tprec_nl} and \ref{eq:tcomm_nl} are sensitive functions of $\rin$, this may drastically change both the number of predicted breaks and the location of any breaking radii.

%%%%%%%%%%%%%%%%%%%%%%%%%%%%%%%%%%%%%%%%%%%%%%%%%%

%%%%%%%%%%%%%%%%%%% CONCLUSION %%%%%%%%%%%%%%%%%%%

%%%%%%%%%%%%%%%%%%%%%%%%%%%%%%%%%%%%%%%%%%%%%%%%%%
\section{Conclusions}
\label{sec:conclusion}

We have used grid-based hydrodynamic simulations to study the behavior of warped discs in the context of circumbinary discs created during early star formation.  We find that the previously known behaviors for warped discs are reproduced well with the grid-based code, and that disc breaking is readily achieved for discs satisfying $\alpha \lesssim h/r$.  From this we propose a viscous criterion for disc breaking, where the disc must undergo significant precession before aligning perpendicular to the precession axis in order to break.

We also derive new formulations of the disc timescales in order to predict the location of a disc break.  Our formulation for disc breaking suggests that breaking events are more likely when the disc is thinner, the inner cavity is smaller, and the disc power-law profiles are steeper.  These criteria are well supported by our simulation results, and the predicted location for the breaking radius is in agreement with the breaks observed in our simulations.  We also show that repeated disc breaking, previously seen in simulations of AGN discs, is predicted by our analytic formulae, though it is unlikely to occur in a protoplanetary disc.  When compared against our simulations, the analytic equations produce breaking radii that are consistent with the observed locations.  However, the sensitivity of these equations makes it difficult to use them in a predictive manner and precisely determine the location of the breaking radius.  We have also applied our breaking formulation against the GW Orionis system to explain the discrepancies between previous simulations and their observed breaks, but better observational constraints are required to accurately assess whether the disc will break on its own.  

We have explored only a small area of the total parameter space for disc breaking in this paper.  Our simulations primarily focus on discs around moderately eccentric $(e \sim 0.5)$ binaries, with equal or moderate mass ratios.  Many characteristics of the overall parameter space, such as the regions where a break forms in the middle of the disc (Section \ref{sec:timescales_gwori}), have yet to be constrained.  In the future, well-constrained disc parameters may be combined with our formulation to determine the breaking radii of warped discs. 

%predicted by our formulations, is possible\textbf{, though unlikely to occur in a protoplanetary disk.}.

%We propose a viscous criterion for disk breaking, as well as a new formulation to predict the location of a disk break.

\section*{Acknowledgements}
We thank our reviewer Alison Young for pointing out the different values for the misalignment angle in the GW Orionis system, and for the many suggestions which helped improve the text of this manuscript.  The simulations in this work were conducted using the NASA High-End Computing (HEC) Program through the NASA Advanced Supercomputing (NAS) Division at Ames Research Center.  This material is based upon work supported in part by the National Aeronautics and Space Administration under Grant No. 80NSSC20M0043 issued through the Nevada NASA Space Grant Consortium.  Z. Z. acknowledges support from NASA award 80NSSC22K1413. Figures in this paper were made with the help of Matplotlib \citep{Hunter2007}, NumPy \citep{Harris2020}, and VisIt \citep{HPV:VisIt}.  RGM and SHL acknowledge support from NASA through grants 80NSSC21K0395 and 80NSSC19K0443.

%%%%%%%%%%%%%%%%%%%%%%%%%%%%%%%%%%%%%%%%%%%%%%%%%%
\section*{Data Availability}
The data used in this paper is available upon request to the corresponding author.



\bibliographystyle{mnras}
\bibliography{ref} % if your bibtex file is called example.bib

%\begin{thebibliography}{99}
%\end{thebibliography}

%%%%%%%%%%%%%%%%%%%%%%%%%%%%%%%%%%%%%%%%%%%%%%%%%%

%%%%%%%%%%%%%%%%% APPENDICES %%%%%%%%%%%%%%%%%%%%%

%\appendix



%%%%%%%%%%%%%%%%%%%%%%%%%%%%%%%%%%%%%%%%%%%%%%%%%%

% Don't change these lines
\bsp	% typesetting comment
\label{lastpage}
\end{document}